\documentclass[journal]{IEEEtran}

\usepackage[fleqn]{amsmath}
\usepackage{amssymb}
\usepackage{float}
\usepackage{graphics}
\usepackage{graphicx}
\usepackage{epsfig}
\usepackage{fancyhdr}
\usepackage{latexsym}
\usepackage{color}
\usepackage{multirow}
\usepackage{amsfonts}
\usepackage{subfigure}
\usepackage{enumerate}

\begin{document}

\title{An ensemble Kushner-Stratonovich-Poisson filter for recursive estimation in nonlinear dynamical systems}

\author{Mamatha~Venugopal,
        Ram~Mohan~Vasu,
        and~Debasish~Roy 

\thanks{M. Venugopal and R. M. Vasu are with the Optical Tomography Lab, Department of Instrumentation and Applied Physics, Indian Institute of Science, Bangalore - 560012, India. email: mamathavenugopal@iap.iisc.ernet.in, vasu@isu.iisc.ernet.in.}
\thanks{D. Roy is with the Computational Mechanics Lab, Department of Civil Engineering, Indian Institute of Science, Bangalore - 560012, India. email: royd@civil.iisc.ernet.in.}
\thanks{Manuscript received Jul XX, 2014; revised XXXX XX, 201X.}}

\markboth{IEEE TRANSACTIONS ON AUTOMATIC CONTROL,~Vol.~XX, No.~XX, XXXX~201X}%
{Shell \MakeLowercase{\textit{et al.}}: Title}

\maketitle

\begin{abstract}

Despite the numerous applications that may be expeditiously modelled by counting processes, stochastic filtering strategies involving Poisson-type observations still remain somewhat poorly developed. In this work, we propose a Monte Carlo stochastic filter for recursive estimation in the context of linear/nonlinear dynamical systems with Poisson-type measurements. A key aspect of the present development is the filter-update scheme, derived from an ensemble approximation of the time-discretized nonlinear filtering equation, modified to account for Poisson-type measurements. Specifically, the additive update through a gain-like correction term, empirically approximated from the innovation integral in the filtering equation, eliminates the problem of particle collapse encountered in many conventional particle filters. Through a few numerical demonstrations, the versatility of the proposed filter is brought forth, first with application to filtering problems with diffusive or Poisson-type measurements and then to an automatic control problem wherein the extremization of the associated cost functional is achieved simply by an appropriate redefinition of the innovation process.

\end{abstract}

\begin{IEEEkeywords}

Automatic control, Bayes method, Filtering, Monte Carlo methods, Poisson processes, Recursive estimation.

\end{IEEEkeywords}

\IEEEpeerreviewmaketitle

\section{Introduction}
\IEEEPARstart{S}{tochastic} filters aim at estimating the (hidden) states of a dynamical system based on a few partial and potentially noisy observations (or measurements) of the system. Stochastic filtering has gained popularity over the recent years owing to their applicability in solving a large class of problems that range from signal processing, control and radar tracking to atmospheric data assimilation. Despite the striking prospects, the Kalman filter (KF) \cite{kalman} remains to date the sole filter that provides closed form analytical expressions for estimating the states/parameters referred to as the process variables in filtering. The KF is a recursive minimum mean square estimator that provides optimal estimates under the assumptions of drift linearity in process and measurement equations, additive Gaussian noises and negligible model uncertainties. However most practical applications of interest are described accurately by non-linear process/measurement models. Attempts at extending the theory of Kalman filtering led to the development of the Extended Kalman filter (EKF) \cite{j}. The EKF provides non-optimal estimates to the process variables via suitable linearizations of the underlying non-linear models. Notwithstanding the numerous applications of the KF and its variants (see \cite{s} and the references therein), these filters typically require elaborate tuning of the process/measurement noise covariances so as to stabilise the simulation models. Moreover, the calculation of Jacobians in highly non-linear large dimensional systems could be tedious and erroneous.

\par The sequential Monte Carlo (SMC) methods \cite{dga}-\cite{amgc}, on the other hand, are well-equipped to handle non-linear and non-Gaussian models, albeit at the cost of more computational power. A realization of the states at a given time is called a particle within a Monte Carlo (MC) set-up. The SMC-based filters, popularly known as the particle filters (PFs), employ Bayesian algorithms to empirically describe the target conditional densities/distributions via an ensemble, which is a finite set of weighted particles. As the size of the ensemble increases, the above set of realizations provide an equivalent representation of the density/distribution with the error reducing at a rate proportional to \(\frac{1}{{\sqrt {{n_e}} }}\) \cite{c}, \({n_e}\) being the ensemble size. Although with the advancements in computing, the implementation of PFs is now numerically feasible, weight-based filters are typically cursed with the problem of sample degeneracy \cite{amgc}, i.e. the sequential collapse of the weights to a point mass. Indeed, the ensemble size required to prevent filter divergence through sample degeneracy increases exponentially with the system dimension as is shown in \cite{sbba}. Although numerous techniques such as the Markov chain Monte Carlo (MCMC) \cite{gt} and implicit sampling \cite{ct} are developed to improve the SMC-based schemes, these methods still remain computationally infeasible in solving large dimensional problems.

\par The evolution of the target distribution of the states conditioned on the measurement history is described by the Zakai equation \cite{z}-\cite{kallianpur} or the normalized Kushner-Stratonovich (KS) equation \cite{kushner}-\cite{b} in non-linear filtering. Both the Zakai and the KS equations are derived sequentially following an It\^{o}'s expansion of the Kallianpur-Striebel formula \cite{ks}, to arrive at an additive update to the evolving conditional distribution as opposed to the usual multiplicative one. There have been many attempts at solving the linear Zakai equation in literature \cite{bk1}-\cite{gppp}. However, the numerical solution of the equation is fraught with serious deficiencies \cite{ir} which are bypassed by its non-linear counterpart, the KS equation. Nevertheless, closed form analytical solutions to the KS equation exist only in cases of linear, Gaussian models \cite{kb}. In order to account for more general cases, approximate MC-based methods that are convergent in distribution to the solution of the KS equation have been developed in \cite{cl}-\cite{cdml}. These methods employ weighted/branching particle systems for filtering and are therefore not free from the scourge of particle collapse. On the other hand, MC-based filters that incorporate the additive update obtainable from the KS equation were formulated and applied successfully in \cite{scvvr}, \cite{rrv}. An additive update attempts at `healing' every particle of the ensemble through a gain-like term that drives the particle towards reducing its corresponding measurement-prediction misfit. Thus, the additive gain-based updates perform the functionality of the multiplicative weights but without the effective reduction in the ensemble size, witnessed in typical SMC-based filters. 

\par Interestingly, all the filters discussed so far assume continuously varying measurements as inputs. However, in many applications that range from packet traffic analysis on computer networks and modelling of communication systems to filtering of biomedical signals, the measurements may simply be a sequence of discrete events in time. The rate of occurrence of these possibly identical events usually contains information on important parameters of interest. For instance, the rate at which photons strike a photodetector depends on the intensity of light. In such cases, it is required to estimate the desired parameters that modulate the rate of the observed events, which are modelled as doubly stochastic Poisson processes in literature \cite{snyder}-\cite{mke}. Although there have been many attempts at devising filtering schemes based on Poisson measurements \cite{snyder}-\cite{mje}, all of them were designed for specific and limited applications. To the best of the authors' knowledge, a full-blown stochastic filtering scheme based on Poisson measurements applicable to a general class of filtering problems has not yet been formulated.

\par In this paper, we propose a Monte Carlo filter based on the KS equation, that is appropriately derived to account for Poisson measurements. As will be shown in the sections to follow, the filtered estimate gives the evolving conditional distribution of the (hidden) states that modulate the instantaneous rate of the observed Poisson process. By employing a gain-based additive particle-wise update, obtained directly from a time-discretized ensemble approximation of the modified KS equation, the proposed filter is free of the weight collapse encountered in most variants of Bayesian filtering. Through illustrative examples, we demonstrate the competence of the filter in estimating the states/ parameters of multi-dimensional dynamical systems, wherein only a few states contribute to the rate modulation of the observed Poisson process. This may be contrasted with the existing schemes in which the filter estimation is limited to the Poisson rate or at most a scalar state variable. In addition, we show that the same filtering scheme may also be applied to solve the conventional filtering problems involving diffusion-type measurements. Thus, we present a unified filter, henceforth referred to as an ensemble KS-Poisson filter (EKSPF) that may be applied for the recursive estimation of diffusion processes in the dual cases of diffusion and/or Poisson-type measurements.

\par The rest of the paper is organized as follows. Section II gives a brief introduction to the filtering problem in general followed by an outline of the derivation of the KS equation for Poisson measurements. The detailed formulation of the proposed filter for Poisson-type and diffusion-type measurements is given in Section III. Pseudo-codes for both the cases are also included for clarity. Section IV demonstrates the performance of the proposed filter through numerical examples and Section V presents the concluding remarks.

\section{Problem Statement and Model Formulation}

\par Within a complete probability space \cite{kallianpur}, say, \(\left( {\Omega ,\mathcal{F},{\rm{P}}} \right)\), equipped with the filtration \({\{ {\mathcal{F}_t}\} _{t \ge 0}}\) such that \({\mathcal{F}_s} \subseteq {\mathcal{F}_t} \subseteq \mathcal{F}\) for \(s \le t\), the stochastic differential equation (SDE) describing the evolution of the process may be written as 
\begin{equation}
d{{\mathbf{X}}_t} = {\boldsymbol{\mu }}\left( {t,{{\mathbf{X}}_t}} \right)dt + {\boldsymbol{\Sigma }}\left( {t,{{\mathbf{X}}_t}} \right)d{{\mathbf{B}}_t}
\label{eq1}
\end{equation}
Here, \({{\bf{X}}_t}: = {\bf{X}}\left( t \right) \in {\mathbb{R}^{{n_x}}}\) is the process vector, \({\boldsymbol{\mu }}:{\mathbb{R}_ + } \times {\mathbb{R}^{{n_x}}} \mapsto {\mathbb{R}^{{n_x}}}\) the nonlinear drift function, \({\boldsymbol{\Sigma }}:{\mathbb{R}_ + } \times {\mathbb{R}^{{n_x}}} \mapsto {\mathbb{R}^{{n_x} \times {n_b}}}\) is the diffusion matrix and \({{\bf{B}}_t}: = {\bf{B}}\left( t \right) \in {\mathbb{R}^{{n_b}}}\) is an \({n_b}\)-dimensional standard \(\rm{P}\)-Brownian motion. The measurement SDE may be written as 
\begin{equation}
d{{\mathbf{Y}}_t} = {\boldsymbol{\lambda }}\left( {t,{{\mathbf{X}}_t}} \right)dt + d{{\mathbf{N}}_t}
\label{eq2}
\end{equation} 
where \({{\bf{Y}}_t} \in {\mathbb{R}^{{n_y}}}\) is a counting process (see \cite{klebaner} for a formal definition of counting/Poisson processes) with bounded non-negative intensity (or rate) \(\boldsymbol{\lambda }\) and \({{\bf{N}}_t}\) a Poisson martingale. Note that \(\boldsymbol{\lambda }\) is a Borel measurable function with the stochastic process \({{\bf{X}}_t}\) as one of its arguments and hence \({{\bf{Y}}_t}\) is referred to as `a doubly stochastic' Poisson process. The existence of solution to SDE (\ref{eq2}) is guaranteed by the Doob-Meyer decomposition of semi-martingales \cite{klebaner}. It is assumed that \({{\bf{X}}_t}\) and \({{\bf{Y}}_t}\) are \({{\mathcal{F}}_t}\)-adapted processes and the coefficients \({\boldsymbol{\mu }}\), \({\boldsymbol{\Sigma }}\) and  \({\boldsymbol{\lambda }}\) are bounded and Lipschitz and further that the necessary conditions for the existence of strong solutions to SDEs (\ref{eq1}) and (\ref{eq2}) hold. Denoting by \({\mathcal{F}}_t^y\) the sub-filtration generated by \({{\bf{Y}}_t}\), the aim of filtering is to determine the measure-valued filtered conditional estimate, \({\pi _t}\left( \phi  \right): = {{\rm{E}}_{\rm{P}}}\left[ {\phi \left( {{{\bf{X}}_t}} \right)|{\mathcal{F}}_t^y} \right]\) of a scalar-valued function \(\phi  \in \mathbb{C}_b^2\) (bounded and twice continuously differentiable) of \({{\bf{X}}_t}\). Here, \({{\rm{E}}_{\rm{P}}}\left[ . \right]\) denotes the expectation operator with respect to the measure \(\rm{P}\).

\par In the theoretical formulations and derivations to follow, we assume \({n_x} = {n_y} = {n_b} = 1\) for clarity of exposition (and the non-bold symbols in the rest of the section indicate scalar-valued processes). Consistent with the conventional filtering schemes, in the present context of Poisson-type measurements, the update step must be so devised as to ensure that the innovation process \({{{Y}}_t} - \int\limits_0^t {{{\lambda}} \left( {s,{X_s}} \right)ds} \) is driven to a \(\left( {\rm{P},{{\mathcal{F}}_t}} \right)\)-Poisson martingale. This is accomplished through a Girsanov change of measures for counting processes that transforms the observed \(\rm{P}\)-Poisson process to a unit intensity Poisson process with respect to a new measure, say \(\rm{Q}\). Specifically, we have the following theorem:
\\
\newtheorem{theorem}{Theorem}
\begin{theorem}
Let \({{Y}_t}\) be an \({{\mathcal{F}}_t}\)-adapted unit Poisson process on \(\left( {\Omega ,\mathcal{F},{\rm{Q}}} \right)\). If \(\lambda\) is an \({{\mathcal{F}}_t}\)-adapted non-negative predictable process such that \(\int\limits_0^t {\lambda \left( {s,{X_s}} \right)ds < \infty }\), for \(t \ge 0\), then 
\begin{equation}
{Z_t} = \exp \left( {\int\limits_0^t {\ln \lambda \left( {s,{{X}_s}} \right)d{Y_s}}  - \int\limits_0^t {\left( {\lambda \left( {s,{X_s}} \right) - 1} \right)ds} } \right)
\label{eq3}
\end{equation}
is a \(\rm{Q}\)-martingale. Furthermore, if for \(t \ge 0\), \({{\rm{E}}_{\rm{Q}}}\left[ {{Z_t}} \right] = 1\) and we define \(d{\rm{P}_t} = {Z_t}d{\rm{Q}_t}\) on \({{\mathcal{F}}_t}\), then \(\rm{P} \ll \rm{Q}\) and \({Y_t} - \int\limits_0^t {\lambda \left( {s,{X_s}} \right)ds} \) is a \(\rm{P}\)-martingale.
\end{theorem}
\begin{IEEEproof}
See \cite{klebaner}.
\end{IEEEproof}

Note here that the Radon-Nikodym derivative \({Z_t}\) is the likelihood ratio which when multiplied to the predictions form the filtered estimate. Denoting the unnormalized filtered estimate as \({\sigma _t}\left( \phi  \right): = {{\rm{E}}_{\rm{Q}}}\left[ {\phi \left( {{{\bf{X}}_t}} \right){Z_t}|{\mathcal{F}}_t^y} \right]\), the filtered estimate may be written as (Kallianpur-Striebel formula)
\begin{equation}
{\pi _t}\left( \phi  \right) = \frac{{{\sigma _t}\left( \phi  \right)}}{{{\sigma _t}\left( 1 \right)}}
\label{eq4}
\end{equation}
However, a multiplicative update, as noted earlier suffers from sample degeneracy especially in higher dimensional filtering problems. Hence, an It\^{o}'s expansion of the multiplicative update is effected to arrive at a nonlinear gain-like correction term that may simply be added to the predictions to obtain the corresponding updates. 
\\
\newtheorem{lm1}{Lemma}
\begin{lm1}
The unnormalized filtered estimate is given by
\begin{equation}
{\sigma _t}\left( \phi  \right) = {\sigma _0}\left( \phi  \right) + \int\limits_0^t {{\sigma _s}\left( {L\phi } \right)ds}  + \int\limits_0^t {{\sigma _{s - }}\left( {\phi \left( {{\lambda _s} - 1} \right)} \right)\left( {d{Y_s} - ds} \right)} 
\label{eq5}
\end{equation}
where \(\left( {L\phi } \right)\left( x \right) = \mu \left( {t,x} \right)\frac{{\partial \phi }}{{\partial x}} + \frac{1}{2}{\Sigma ^2}\left( {t,x} \right)\frac{{{\partial ^2}\phi }}{{\partial {x^2}}}\) and \({\lambda _s}: = \lambda \left( {s,{X_s}} \right)\).
\end{lm1}
\begin{IEEEproof}
Since \({Z_t}\) is the stochastic exponential of \(\int\limits_0^t {\left( {{\lambda _s} - 1} \right)\left( {d{Y_s} - ds} \right)} \), we have \[d{Z_t} = {Z_{t - }}\left( {{\lambda _t} - 1} \right)\left( {d{Y_t} - dt} \right)\]
 Now,
\(d\left( {\phi \left( {{X_t}} \right){Z_t}} \right) = {Z_t}L\left( {\phi \left( {{X_t}} \right)} \right)dt + {Z_t}\phi '\left( {{X_t}} \right)d{B_t} \\
\,\,\,\,\,\,\,\,\,\,\,\,\,\,+ \phi \left( {{X_t}} \right){Z_{t - }}\left( {{\lambda _t} - 1} \right)\left( {d{Y_t} - dt} \right)\)
\\Applying the conditional expectation operator, \({{\rm{E}}_{\rm{Q}}}\left[ {.|{\mathcal{F}}_t^y} \right]\) to the integral form of the above expression and noting that \({{\rm{E}}_{\rm{Q}}}\left[ {\int\limits_0^t {{Z_s}\phi '\left( {{X_s}} \right)d{B_s}|{\mathcal{F}}_t^y} } \right] = 0\) leads to (\ref{eq5}).
\end{IEEEproof}

Using the expression for \({\sigma _t}\left( \phi  \right)\) in the Kallianpur-Striebel formula yields the normalized filtered estimate. The following theorem gives the expression for the evolving conditional distribution of the states conditioned on the Poisson observation history.\\
\begin{theorem}
(Kushner-Stratonovich-Poisson equation)
\begin{IEEEeqnarray}{c}
{\pi _t}\left( \phi  \right) = {\pi _0}\left( \phi  \right) + \int\limits_0^t {{\pi _s}\left( {L\phi } \right)ds} + \IEEEnonumber*\\  
\int\limits_0^t {\left\{ {\frac{{{\pi _{s - }}\left( {\lambda \phi } \right) - {\pi _{s - }}\left( \phi  \right){\pi _{s - }}\left( \lambda  \right)}}{{{\pi _{s - }}\left( \lambda  \right)}}} \right\}\left( {d{Y_s} - {\pi _{s - }}\left( \lambda  \right)ds} \right)} \IEEEyesnumber*
\label{eq6}
\end{IEEEeqnarray}
\end{theorem}
\begin{IEEEproof}
See Appendix A.
\end{IEEEproof}

In the next section, the Kushner-Stratonovich-Poisson (KSP) equation is written in a more general form to accommodate vector-valued processes and measurements. The filtering algorithm is then put forth and elaborated upon through the prediction-update steps, first for Poisson-type and then for diffusion-type measurements.

\section{Filtering Strategy}
Given a partition \(0 = {t_0} < {t_1}... < {t_{{n_t}}} = T\) of the time interval of interest, a time-discretized version of the generalized KSP equation is given by:
\begin{IEEEeqnarray}{c}
{\pi _t}\left( \phi  \right) = {\pi _{{t_i}}}\left( \phi  \right) + \int\limits_{{t_i}}^t {{\pi _s}\left( {L\phi } \right)ds} + \IEEEnonumber*\\
\sum\limits_{l = 1}^{{n_y}} {\int\limits_{{t_i}}^t {\left\{ {\frac{{{\pi _{s - }}\left( {{\lambda ^l}\phi } \right) - {\pi _{s - }}\left( \phi  \right){\pi _{s - }}\left( {{\lambda ^l}} \right)}}{{{\pi _{s - }}\left( {{\lambda ^l}} \right)}}} \right\}\left( {dY_s^l - {\pi _{s - }}\left( {{\lambda ^l}} \right)ds} \right)} } \IEEEnonumber*\\ \IEEEyesnumber\IEEEeqnarraynumspace
\label{eq7}
\end{IEEEeqnarray}
for \(t \in ({t_i},{t_{i + 1}}]\), where \({{\bf{Y}}_t} = \left( {Y_t^1,...,Y_t^{{n_y}}} \right)\) and \({{\boldsymbol{\lambda }}_t} = \left( {\lambda _t^1,...,\lambda _t^{{n_y}}} \right)\). Moreover, \[\left( {L\phi } \right)\left( {\bf{x}} \right) = \sum\limits_{l = 1}^{{n_x}} {{\mu ^l}\left( {t,{\bf{x}}} \right)\frac{{\partial \phi \left( {\bf{x}} \right)}}{{\partial {x^l}}}}  + \frac{1}{2}\sum\limits_{l = 1}^{{n_x}} {\sum\limits_{m = 1}^{{n_x}} {{{\boldsymbol{\Xi }}_{lm}}\left( {t,{\bf{x}}} \right)\frac{{{\partial ^2}\phi \left( {\bf{x}} \right)}}{{\partial {x^l}\partial {x^m}}}} } \] where \({\bf{x}} = \left( {{x^1},...,{x^{{n_x}}}} \right)\), \({\mu ^l}\) denotes the \({l^{th}}\) component of the vector \({\boldsymbol{\mu }}\) and \({{\bf{\Xi }}_{lm}}\) the \({\left( {l,m} \right)^{th}}\) element of the matrix \({\boldsymbol{\Xi }} = {\boldsymbol{\Sigma }}{{\boldsymbol{\Sigma }}^T}\).  Owing to the inherent difficulties involved in the evaluation of the integrals appearing in the KSP equation, a further simplification is effected through an Euler-Maruyama (EM) approximation as follows:
\begin{IEEEeqnarray}{c}
{\pi _{t_{i+1}}}\left( \phi  \right) = {\pi _{{t_i}}}\left( \phi  \right) + {\pi _{{t_i}}}\left( {L\phi } \right)\Delta t + \IEEEnonumber*\\
 \sum\limits_{l = 1}^{{n_y}} {\left\{ {\frac{{{\pi _{{t_i}}}\left( {{\lambda ^l}\phi } \right) - {\pi _{{t_i}}}\left( \phi  \right){\pi _{{t_i}}}\left( {{\lambda ^l}} \right)}}{{{\pi _{{t_i}}}\left( {{\lambda ^l}} \right)}}} \right\}\left( {\Delta Y_{{t_i}}^l - {\pi _{{t_i}}}\left( {{\lambda ^l}} \right)\Delta t} \right)} \IEEEnonumber*\\ \IEEEyesnumber\IEEEeqnarraynumspace
 \label{eq8}
\end{IEEEeqnarray}
where \(\Delta t = {t_{k + 1}} - {t_k}\) is assumed to be uniform for \(k = 0,...,{n_t} - 1\) and \(\Delta Y_{{t_i}}^l = Y_{{t_{i + 1}}}^l - Y_{{t_i}}^l\). Note that \({{\bf{Y}}_0} = \bf{0}\).

\subsection{Filtering scheme for Poisson-type measurements}
 The measurement equation applicable here would be given by (\ref{eq2}), wherein the actual measurement is a non-homogeneous point process whose rate is instantaneously modulated by the (hidden) states that need to be estimated. Once again, for expositional clarity, we consider \(\phi \left( {\bf{x}} \right) = {\bf{x}}\). As is evident from (\ref{eq8}), the first two terms on the RHS correspond to the prediction step with the last available solution at $t_i$ serving as the initial condition. In the MC set-up that we adopt here, the filtered estimate in (\ref{eq8}) is arrived at empirically via an ensemble of independent realizations of the state. Assuming that \({n_e}\) denotes the ensemble size, let \(\left\{ {{{{\bf{\hat X}}}_{{t_i}}}\left( j \right)} \right\}_{j = 1}^{{n_e}}\) be the updated set of particles at time \({t_i}\). Then, the filter algorithm obtaining the solution at the time instant \({t_i+1}\) consists of the following two steps.\\
\\{\it{Prediction}}: The prediction equation is obtained from the process SDE (\ref{eq1}) through a time-discretized EM approximation (or a higher order integrator for SDEs if available), which may be used to propagate each particle separately as follows:
\begin{IEEEeqnarray}{rCl}
{{\mathbf{\tilde X}}_{{t_{i + 1}}}}\left( j \right) = {{\mathbf{\hat X}}_{{t_i}}}\left( j \right) + {\boldsymbol{\mu }}\left( {{t_{i + 1}},{{{\mathbf{\hat X}}}_{{t_i}}}\left( j \right)} \right)\Delta t \IEEEnonumber* \\
 + {\boldsymbol{\Sigma }}\left( {{t_{i + 1}},{{{\mathbf{\hat X}}}_{{t_i}}}\left( j \right)} \right)\Delta {{\mathbf{B}}_{{t_i}}}\left( j \right), 
j = 1,...,{n_e} \IEEEyesnumber*
\label{eq9}
\end{IEEEeqnarray}
where the superscript tilde denotes prediction and \(\Delta {{\bf{B}}_{{t_i}}}\left( j \right) = {{\bf{B}}_{{t_{i + 1}}}}\left( j \right) - {{\bf{B}}_{{t_i}}}\left( j \right)\).
\\
\\{\it{Update}}: The particle-wise update obtainable from the third term in (\ref{eq8}) may be written as
\begin{IEEEeqnarray}{c}
\hat X_{{t_{i + 1}}}^m\left( j \right) = \tilde X_{{t_{i + 1}}}^m\left( j \right) + \IEEEnonumber* \\
\sum\limits_{l = 1}^{{n_y}} {\left( {\begin{array}{ccccccccccccccc}
  {\left( {\frac{{{{\tilde \pi }_{{t_{i + 1}}}}\left( {{X^m}{\lambda ^l}\left( {{{{\mathbf{\tilde X}}}_{{t_{i + 1}}}}} \right)} \right) - {{\tilde \pi }_{{t_{i + 1}}}}\left( {{X^m}} \right){{\tilde \pi }_{{t_{i + 1}}}}\left( {{\lambda ^l}\left( {{{{\mathbf{\tilde X}}}_{{t_{i + 1}}}}} \right)} \right)}}{{{{\tilde \pi }_{{t_{i + 1}}}}\left( {{\lambda ^l}\left( {{{{\mathbf{\tilde X}}}_{{t_{i + 1}}}}} \right)} \right)}}} \right) \times } \\ 
  {\left( {\Delta Y_{{t_i}}^l - {\lambda ^l}\left( {{{{\mathbf{\tilde X}}}_{{t_{i + 1}}}}\left( j \right)} \right)\Delta t} \right)} 
\end{array}} \right)} , \IEEEnonumber* \\
m = 1,...,{n_x},j = 1,...,{n_e} \IEEEyesnumber*
\label{eq10}
\end{IEEEeqnarray}
where \({\pi _{{t_{i + 1}}}}\left( . \right): = \pi \left( {{{\left( . \right)}_{{t_{i + 1}}}}} \right)\) and, as before, the superscript tilde represents the predicted particles. Specifically, we have \({\tilde \pi _{{t_{i + 1}}}}\left( {{X^l}} \right): = \frac{1}{{{n_e}}}\sum\limits_{k = 1}^{{n_e}} {\tilde X_{{t_{i + 1}}}^l\left( k \right)}\). The step-by-step algorithm is provided below for further clarity. \\
\\{\it{Algorithm 1}}\\
Given a partition of the time interval of interest, say as \(0 = {t_0} < {t_1}... < {t_{{n_t}}} = T\), and the measurements, \(\left\{ {{{\bf{Y}}_{{t_i}}}} \right\}_{i = 1}^{{n_t}}\). Assume that  \(\Delta {t_i} = {t_{i + 1}} - {t_i} = \Delta t, {\text{uniformly for }} i = 0,...,{n_t} - 1\). Note, however that the actual Poisson events may be registered at unequal time intervals based on their respective intensities. Nevertheless, the filtering may be carried out at the above set time intervals. Also, set the ensemble size, \({n_e}\).\\ \\
1) Generate the initial ensemble of states, \(\left\{ {{{{\bf{\hat X}}}_0}\left( j \right)} \right\}_{j =   1}^{{n_e}}\), drawn for instance from a Gaussian distribution of zero-mean and assumed covariance. Set \(i = 0, {t_0} = 0\). \\ \\
2) ({\it{Prediction}}) Obtain the predicted set of particles, \(\left\{ {{{{\bf{\tilde X}}}_{{t_{i + 1}}}}\left( j \right)} \right\}_{j = 1}^{{n_e}}\) for time \({t_{i + 1}} = \left( {i + 1} \right)\Delta t\) using (\ref{eq9}). \\ \\
3) Compute \({\boldsymbol{\lambda }}\left( {{{{\bf{\tilde X}}}_{{t_{i + 1}}}}\left( j \right)} \right): = {\boldsymbol{\lambda }}\left( {{t_{i + 1}},{{{\bf{\tilde X}}}_{{t_{i + 1}}}}\left( j \right)} \right)\) for  \(j = 1,...,{n_e}\). \\ \\
4) Evaluate each column of the gain matrix, \({{\bf{G}}_{{t_{i + 1}}}} = \left[ {G_{{t_{i + 1}}}^1,...,G_{{t_{i + 1}}}^{{n_y}}} \right] \in {\mathbb{R}^{{n_x} \times {n_y}}}\) as
\[G_{{t_{i + 1}}}^l = \left\{ {\frac{{\sum\limits_{k = 1}^{{n_e}} {{\lambda ^l}\left( {{{{\bf{\tilde X}}}_{{t_{i + 1}}}}\left( k \right)} \right){{{\bf{\tilde X}}}_{{t_{i + 1}}}}\left( k \right)} }}{{\sum\limits_{k = 1}^{{n_e}} {{\lambda ^l}\left( {{{{\bf{\tilde X}}}_{{t_{i + 1}}}}\left( k \right)} \right)} }} - \frac{1}{{{n_e}}}\sum\limits_{k = 1}^{{n_e}} {{{{\bf{\tilde X}}}_{{t_{i + 1}}}}\left( k \right)} } \right\}\]
for \(l = 1,...,{n_y}\). \\ \\
5) ({\it{Update}}) Update each particle separately according to the formula
\[{{\bf{\hat X}}_{{t_{i + 1}}}}\left( j \right) = {{\bf{\tilde X}}_{{t_{i + 1}}}}\left( j \right) + {{\bf{G}}_{{t_{i + 1}}}}\left( {\Delta {{\bf{Y}}_{{t_i}}} - {\boldsymbol{\lambda }}\left( {{{{\bf{\tilde X}}}_{{t_{i + 1}}}}\left( j \right)} \right)} \right)\]
for \(j = 1,...,{n_e}\). \\ \\
6) Calculate the state estimate at \({t_{i + 1}}\) as 
\begin{equation}
{{\mathbf{\bar X}}_{{t_{i + 1}}}} = \frac{1}{{{n_e}}}\sum\limits_{k = 1}^{{n_e}} {{{{\mathbf{\hat X}}}_{{t_{i + 1}}}}\left( k \right)} 
\label{eq11}
\end{equation}\\
7) Set \(i \to i + 1\). If \({t_i} = T\), then stop the algorithm; else, go to step 2.

\subsection{Filtering scheme for diffusion-type measurements}
The discrete-time evolution of diffusion-type measurements may be written in an algebraic form as follows
\begin{equation}
{{\mathbf{M}}_{{t_{i + 1}}}} = {\mathcal{M}}\left( {{t_{i + 1}},{{\mathbf{X}}_{{t_{i + 1}}}}} \right) + \Delta {{\mathbf{W}}_{{t_i}}}
\label{eq12}
\end{equation}
where, for \(t \in \left[ {0,T} \right]\), \({{\bf{M}}_t} \in {\mathbb{R}^{{n_m}}}\) denotes the measurement, \({\mathcal{M}}:{\mathbb{R}_ + } \times {\mathbb{R}^{{n_x}}} \to {\mathbb{R}^{{n_m}}}\) the measurement operator, possibly nonlinear and \({{\bf{W}}_t}\) is an \({n_m}\)-dimensional non-standard Brownian motion representing the measurement noise. Also, \(\Delta {{\bf{W}}_{{t_i}}} = {{\bf{W}}_{{t_{i + 1}}}} - {{\bf{W}}_{{t_i}}}\). The process SDE in this case remains the same as (\ref{eq1}). Based on the actual measurements, \(\left\{ {{{\bf{M}}_t}} \right\}\), we generate virtual Poisson measurements whose rates are appropriately defined functions of the measurements. (Measurements obtained from certain transduction elements, e.g. photon current from a photomultiplier tube with shot noise added, comes with an added Poisson-type noise.) The typical rate functions chosen in our numerical examples are of the form
\begin{equation}
{\boldsymbol{\Lambda }}\left( {{{\mathbf{X}}_t}} \right): = {\mathbf{A}}\left| {{{\mathbf{M}}_t}} \right|
\label{eq13}
\end{equation}
where \(\left| . \right|\) is the modulus operator that imposes the non-negativity condition on \({\boldsymbol{\Lambda }}\). Also, \({\bf{A}} = diag\left( {{\alpha _1},...,{\alpha _{{n_m}}}} \right)\) is a diagonal matrix with scalar entries \(\left\{ {{\alpha _k}} \right\}_{k = 1}^{{n_m}}\) that ensure that each of the rate components in the vector, \({\boldsymbol{\Lambda }}\) are sufficiently high to simulate realistic Poisson processes (e.g. the arrival of packets in computer networks). The slight change in the notation of the intensity function above is to highlight the fact that the above expression is employed merely to generate the Poisson measurements. The actual expression for \({\boldsymbol{\lambda }}\) that would be used throughout the filtering steps is given by
\begin{equation}
{\boldsymbol{\lambda }}\left( {{{\mathbf{X}}_t}} \right): = {\mathbf{A}}\left| {{\mathcal{M}}\left( {t,{{\mathbf{X}}_t}} \right)} \right|
\label{eq14}
\end{equation}
 We now generate Poisson processes for the time duration for which the measurements, \(\left\{ {{{\bf{M}}_t}} \right\}\) are available.  The new Poisson-type measurement equation that incorporates the actual measurements in the virtual Poisson intensities may then be given as follows: 
\begin{equation}
d{{\mathbf{Y}}_t} = {\boldsymbol{\lambda }}\left( {{{\mathbf{X}}_t}} \right)dt + d{{\mathbf{N}}_t}
\label{eq15}
\end{equation}
which is the same equation as (\ref{eq2}). Nevertheless, in this case, \({{\bf{Y}}_t} \in {\mathbb{R}^{{n_m}}}\) denotes a virtual Poisson observation constructed from the actual diffusion-type measurements. 

\par The prediction-update steps are essentially the same as in the earlier case of Poisson-measurements and are therefore not repeated here. However, an algorithm is given below to clarify the steps involved in the present filtering scheme. \\
\\{\it{Algorithm 2}}\\
Consider a partition of the time interval of interest based on the availability of measurements, say as 
\(0 = {t_0} < {t_1}... < {t_{{n_t}}} = T\) and the measurements, \(\left\{ {{{\bf{M}}_{{t_i}}}} \right\}_{i = 1}^{{n_t}}\). 
Assume that the time interval between any two successive measurements is a constant, i.e. \(\Delta {t_i} = {t_{i + 1}} - {t_i} = \Delta t,i = 0,...,{n_t} - 1\). Set the ensemble size, \({n_e}\). Choose \(\left( {{\alpha _1},...,{\alpha _{{n_m}}}} \right)\) based on the relative magnitudes of the scalar components of \(\left\{ {{{\bf{M}}_{{t_i}}}} \right\}_{i = 1}^{{n_t}}\). \\ \\
1) Generate the virtual Poisson measurements, \(\left\{ {{{\bf{Y}}_{{t_i}}}} \right\}_{i = 1}^{{n_t}}\) based on the intensity vector, \({\boldsymbol{\Lambda }} = \left( {{\Lambda ^1},...,{\Lambda ^{{n_m}}}} \right)\) calculated as follows:
\[\Lambda _{{t_i}}^l = {\alpha_l}\left| {M_{{t_i}}^l} \right|\]
for \(i = 1,....,{n_t}\) and \(l = 1,...,{n_m}\). Also, \({\bf{M}} = \left( {{M^1},...,{M^{{n_m}}}} \right)\). We take \({{\bf{Y}}_0} = {\bf{0}}\). A pseudo-code to generate the Poisson measurements is given after the algorithm. \\ \\
2) Generate the initial ensemble of states, \(\left\{ {{{{\bf{\hat X}}}_0}\left( j \right)} \right\}_{j =   1}^{{n_e}}\) according to an appropriate distribution, e.g. zero-mean Gaussian with a given covariance. Set \(i = 0,{t_0} = 0\). \\ \\
3) ({\it{Prediction}}) Obtain the predicted set of particles, \(\left\{ {{{{\bf{\tilde X}}}_{{t_{i + 1}}}}\left( j \right)} \right\}_{j = 1}^{{n_e}}\) for time \({t_{i + 1}} = \left( {i + 1} \right)\Delta t\) using (\ref{eq9}). \\ \\
4) Compute \({\boldsymbol{\lambda }}\left( {{{{\bf{\tilde X}}}_{{t_{i + 1}}}}\left( j \right)} \right): = {\bf{A}}\left| {{\mathcal{M}}\left( {{t_{i + 1}},{{{\bf{\tilde X}}}_{{t_{i + 1}}}}\left( j \right)} \right)} \right|\) for \(j = 1,...,{n_e}\). \\ \\
5) Evaluate each column of the gain matrix, \({{\bf{G}}_{{t_{i + 1}}}} = \left[ {G_{{t_{i + 1}}}^1,...,G_{{t_{i + 1}}}^{{n_y}}} \right] \in {\mathbb{R}^{{n_x} \times {n_y}}}\) as
\[G_{{t_{i + 1}}}^l = \left\{ {\frac{{\sum\limits_{k = 1}^{{n_e}} {{\lambda ^l}\left( {{{{\bf{\tilde X}}}_{{t_{i + 1}}}}\left( k \right)} \right){{{\bf{\tilde X}}}_{{t_{i + 1}}}}\left( k \right)} }}{{\sum\limits_{k = 1}^{{n_e}} {{\lambda ^l}\left( {{{{\bf{\tilde X}}}_{{t_{i + 1}}}}\left( k \right)} \right)} }} - \frac{1}{{{n_e}}}\sum\limits_{k = 1}^{{n_e}} {{{{\bf{\tilde X}}}_{{t_{i + 1}}}}\left( k \right)} } \right\}\]
for \(l = 1,...,{n_y}\). \\ \\ 
6) ({\it{Update}}) Update each particle separately according to the formula
\[{{\bf{\hat X}}_{{t_{i + 1}}}}\left( j \right) = {{\bf{\tilde X}}_{{t_{i + 1}}}}\left( j \right) + {{\bf{G}}_{{t_{i + 1}}}}\left( {\Delta {{\bf{Y}}_{{t_i}}} - {\boldsymbol{\lambda }}\left( {{t_{i + 1}},{{{\bf{\tilde X}}}_{{t_{i + 1}}}}\left( j \right)} \right)} \right)\]
for \(j = 1,...,{n_e}\). \\ \\
7) Calculate the state estimate at \({t_{i + 1}}\) according to the formula in (\ref{eq11}). \\ \\
8) Set \(i \to i + 1\). If \({t_i} = T\), then stop the algorithm; else, go to step 3. 

\par Finally, we end this section with a pseudo-code to generate the Poisson measurements in numerical simulations. \\
\\{\it{Pseudo-code to generate a Poisson process}}\\
Assume \({n_y}\)/\({n_m}\)\( = 1\). Set \[i = 0,count = 0,arrival\_time =  - \frac{{\log (rand)}}{{{\lambda _{{t_1}}}}}\]
1) Set \(i \to i + 1\). If \({t_i} = T\), then stop the algorithm; else, go to step 2. \\
2) If \({t_i} <  = arrival\_time\), then \[\begin{array}{l}count = count + 1\\arrival\_time = arrival\_time - \frac{{\log \left( {rand} \right)}}{{{\lambda _{{t_i}}}}}\end{array}\] else go to step 3. \\
3) Set \({Y_{{t_i}}} = count\). Go to step 1.\\
For ease of reference, the proposed filtering approach will sometimes be called the Ensemble Kushner-Stratonovich-Poisson Filter (EKSPF).

\section{Numerical Illustrations}
We consider three different nonlinear estimation problems to numerically demonstrate the performance of the EKSPF. The first problem is on target tracking involving only diffusion-type measurements. The second example involves the state/parameter estimation of a 5-storey shear frame, a mechanical oscillator, under harmonic dynamic loading. This problem is posed as one with strictly Poisson-type measurement noises to assess the filter performance in this case. In both the cases, the numerical results are compared to a particle-wise update version of the unbiased ensemble square root filter \cite{ldn} (henceforth referred to as the PUESRF), which is known to generally outperform the typical SMC-based filters in large-dimensional system identification problems. It must however be noted that a comparison with the PUESRF is strictly valid only for the first example, as the PUESRF can treat only diffusion-type measurements. Nevertheless, in the second problem, the performance comparison of the two filters is reported after allowing for a low-intensity diffusive measurement noise whilst applying the PUESRF. The third example, which involves no explicit measurement equation, is on active structural control wherein the filter algorithm estimates the control force that minimizes the structural responses under external loading. Further details pertaining to the last example are given later. The ensemble size \({n_e}\) is kept at 200 for both the filters in all examples.

\subsection{Target Tracking}
In this problem, we estimate the trajectory of a manoeuvering target from the noisy sensor data (bearing and range). The discrete-time process model of the target may be written as 
\[{{\bf{X}}_{{t_{i + 1}}}} = \Upsilon {{\bf{X}}_{{t_i}}} + \Gamma \left( {{{\bf{a}}_{{t_i}}} + {{\bf{m}}_{{t_i}}}} \right)\]
where \({\bf{X}} = {\left[ {x\,\,\dot x\,\,y\,\,\dot y} \right]^T}\) represents the state vector comprising of the positions and velocities in the {\it{x}}- and {\it{y}}- directions respectively. Also, \(\Upsilon  = \left[ {\begin{array}{*{20}{c}}1&{\Delta t}&0&0\\0&1&0&0\\0&0&1&{\Delta t}\\0&0&0&1\end{array}} \right]\) and \(\Gamma  = \left[ {\begin{array}{*{20}{c}}{\frac{1}{2}{{\left( {\Delta t} \right)}^2}}&0\\{\Delta t}&0\\0&{\frac{1}{2}{{\left( {\Delta t} \right)}^2}}\\0&{\Delta t}\end{array}} \right]\). Moreover, \(\Delta t\) is the sampling time period, \({{\bf{a}}_{{t_i}}} = {\left[ {a_{{t_i}}^x,a_{{t_i}}^y} \right]^T}\) is the random acceleration vector currently characterized  as Brownian and \({{\bf{m}}_{{t_i}}} = {\left[ {m_{{t_i}}^x,m_{{t_i}}^y} \right]^T}\) is non-zero only at the manoeuvering time instants. The sensor is assumed to be at the origin \(\left( {{x_0},{y_0}} \right)\) and the measured bearing and range may be written in a time-discrete form as follows: \[{{\bf{M}}_{{t_{i + 1}}}} = \left( {\begin{array}{*{20}{c}}{{{\tan }^{ - 1}}\left( {\frac{{{y_{{t_{i + 1}}}} - {y_0}}}{{{x_{{t_{i + 1}}}} - {x_0}}}} \right)}\\{\sqrt {{{\left( {{x_{{t_{i + 1}}}} - {x_0}} \right)}^2} + {{\left( {{y_{{t_{i + 1}}}} - {y_0}} \right)}^2}} }\end{array}} \right) + \Delta {{\bf{W}}_{{t_i}}}\] The initial state of the target is at \(\left[ {0.5m,{\rm{ }}3m{s^{ - 1}},{\rm{ }}1m,{\rm{ }}1m{s^{ - 1}}} \right]\) and over the course of its trajectory, it experiences a 4-leg manoeuvering sequence at 20s, 30s, 60s and 80s respectively with accelerations \(\left[ { - 40m{s^{ - 2}},40m{s^{ - 2}}} \right]\), \(\left[ {25m{s^{ - 2}}, - 25m{s^{ - 2}}} \right]\), \(\left[ {25m{s^{ - 2}}, - 25m{s^{ - 2}}} \right]\) and \(\left[ { - 60m{s^{ - 2}},60m{s^{ - 2}}} \right]\). The target moves along straight lines in between the manoeuvers and the trajectory ends at 100s. Here the measurements are generated synthetically by adding 5\% standard Gaussian noise to the bearing and range data constructed from the above trajectory.
\par Virtual Poisson measurements are generated for the EKSPF as per the algorithm in the previous section with \({\alpha _1} = {10^5}\) and \({\alpha _2} = {10^4}\). The process models for both the filters are as given above with the accelerations, \({{\bf{a}}_{{t_i}}}\) and \({{\bf{m}}_{{t_i}}}\) characterized as zero-mean Brownian motions with intensities 0.1 and 100 for \(i = 1,...,{n_t}\).  Fig. 1 shows the trajectories estimated by the EKSPF and the PUESRF when the initial position of the target is known. In spite of an added modulus-type nonlinearity in the generation of the virtual measurements, the EKSPF is seen to closely follow the manoeuvering target. In particular, the EKSPF latches on to the trajectory much faster in comparison to the PUESRF when the tracking is initiated from a faraway point as shown in Fig. 2. A possible reason for this could be the inability of the PUESRF to resolve sources of non-Gaussianity in the data. Figs. 3 and 4 plot the time histories of the root mean square error (RMSE) in the estimation of the {\it{x}}- and {\it{y}}- coordinates respectively for the case in Fig. 2. The RMSE is computed over 100 independent MC runs of the EKSPF and the PUESRF. Fig. 5 is a further demonstration of the ability of the EKSPF in tracking a circular trajectory.   

\begin{figure}[!t]
\centering
\includegraphics[width=3in]{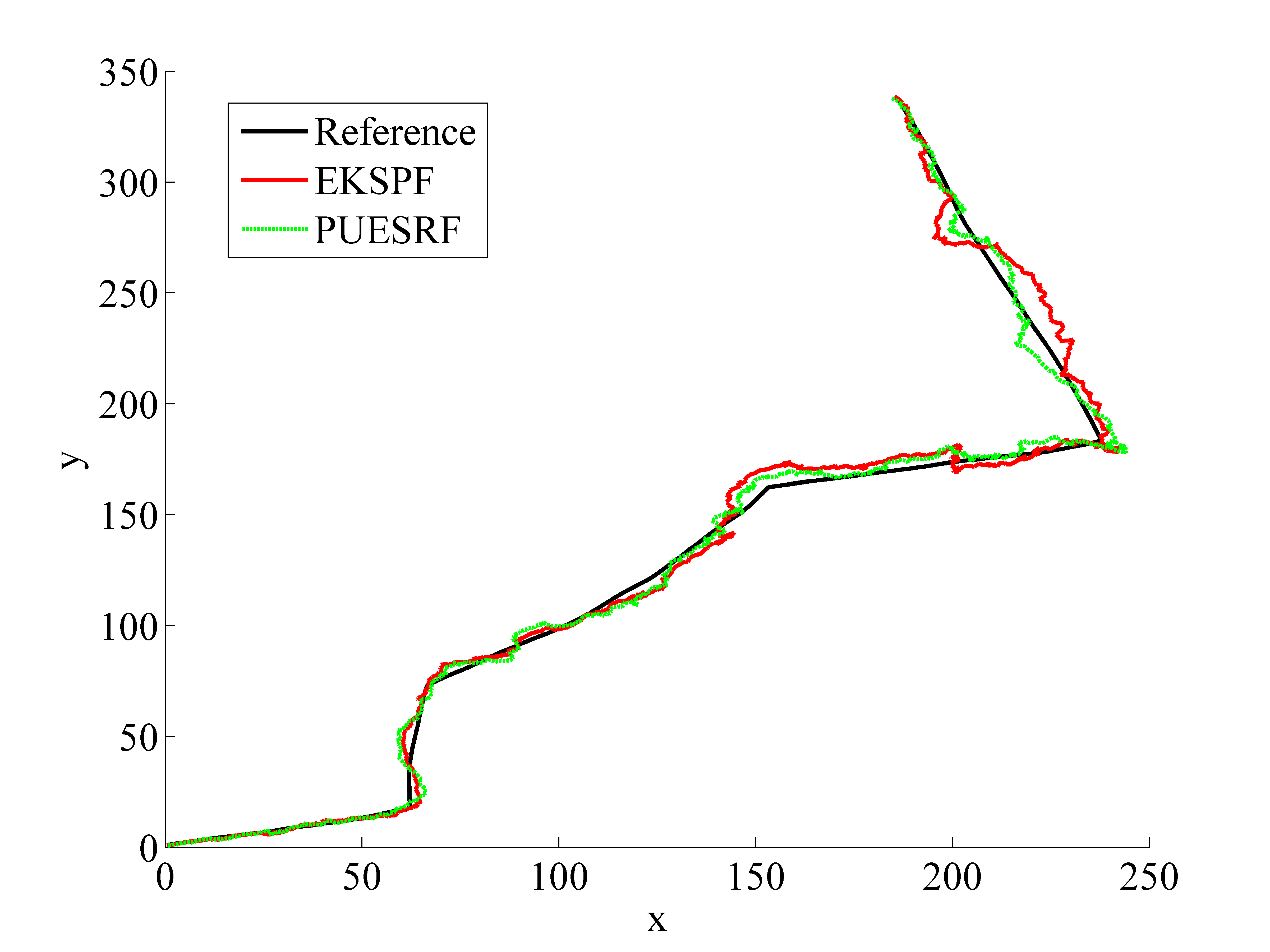}
\caption{The target trajectories in the {\it{x}}-{\it{y}} plane estimated by the EKSPF and the PUESRF against the reference when the initial position of the target is known.}
\label{fig1}
\end{figure}
\begin{figure}[!t]
\centering
\includegraphics[width=3in]{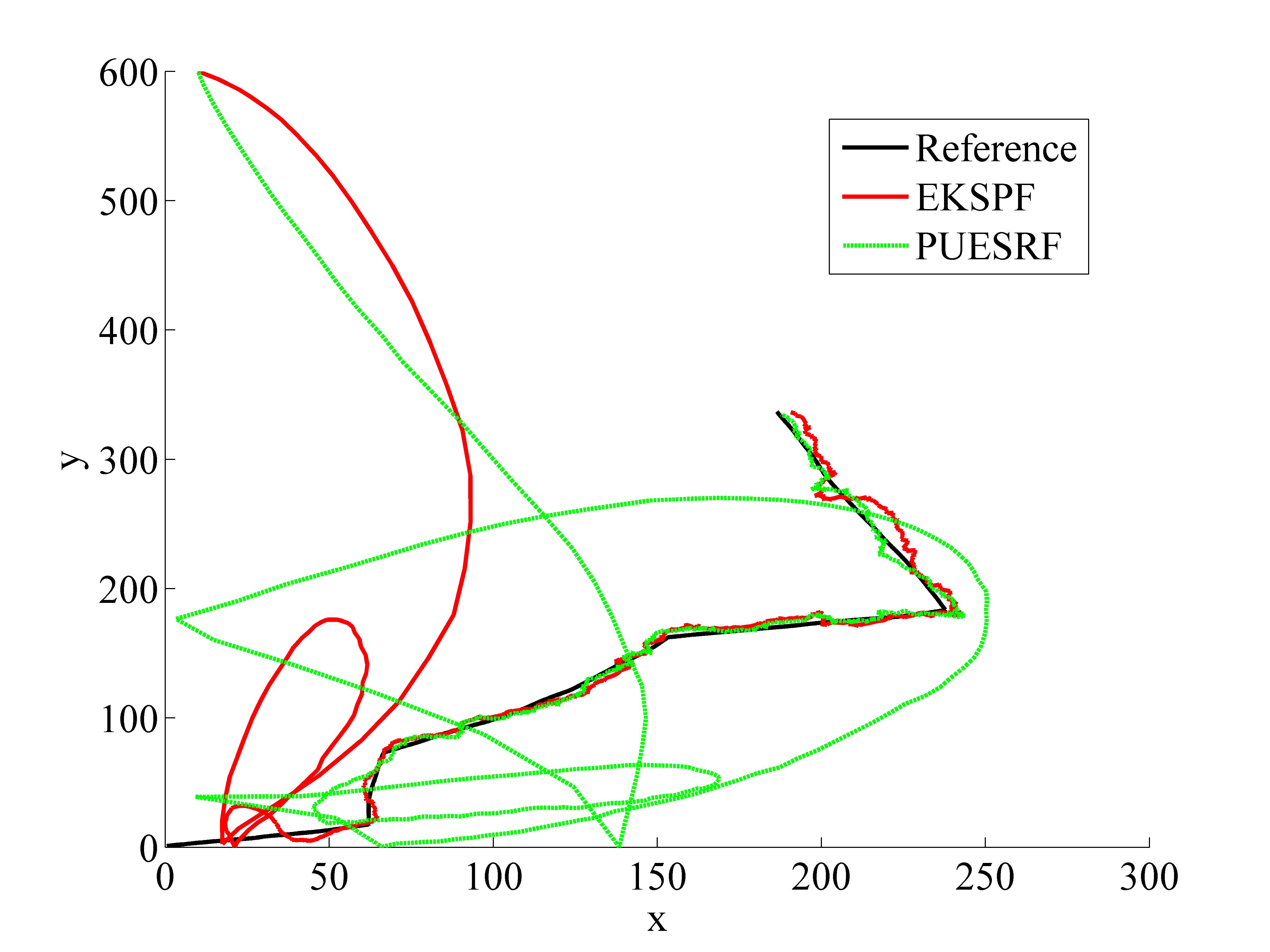}
\caption{The target trajectories in the {\it{x}}-{\it{y}} plane estimated by the EKSPF and the PUESRF against the reference when the initial position of the target is unknown. In this case, the tracking starts from a faraway point.}
\label{fig2}
\end{figure}
\begin{figure}[!t]
\centering
\includegraphics[width=3in]{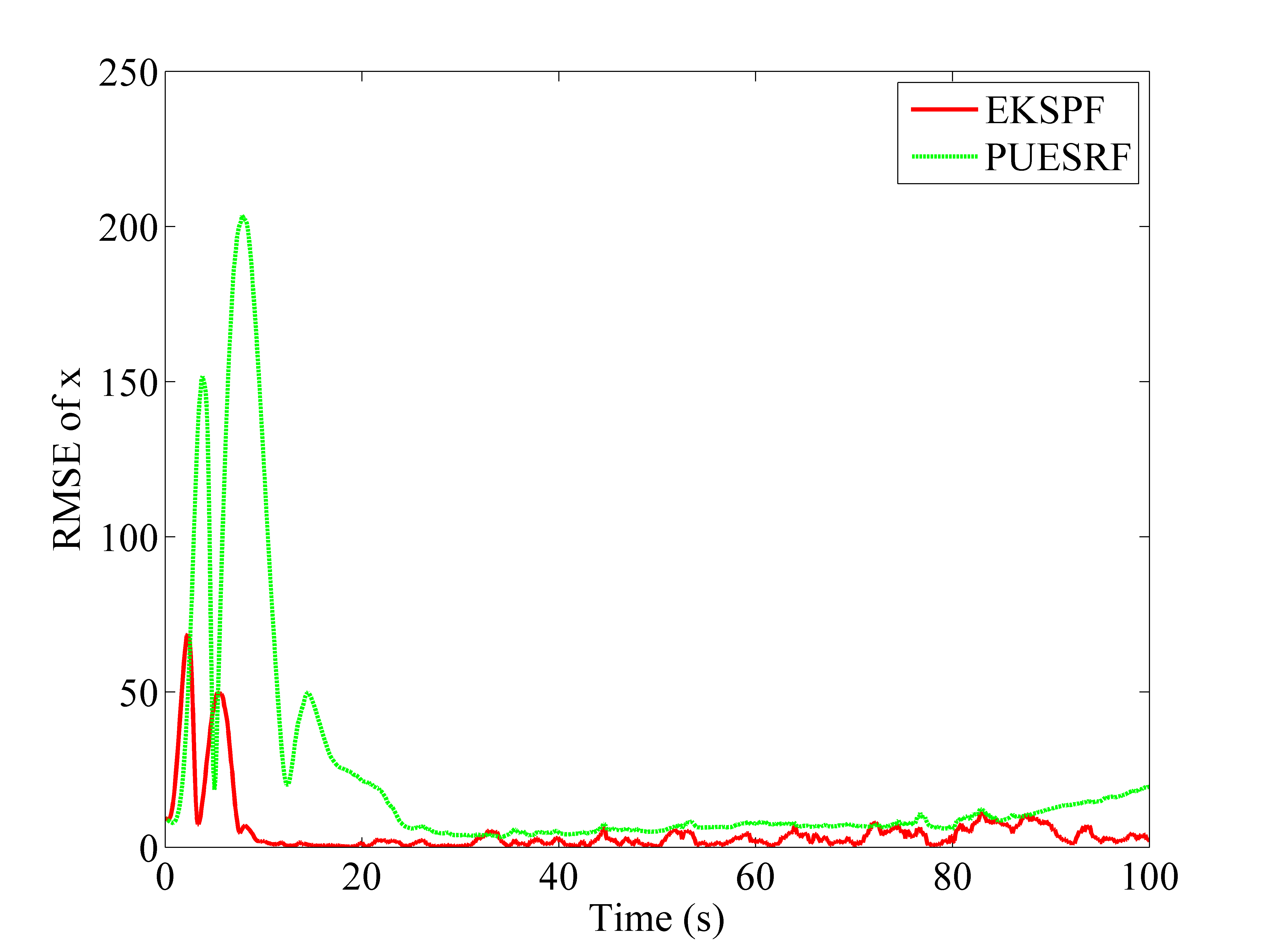}
\caption{RMSE versus time plot of the {\it{x}}-coordinate for the case depicted in Fig. 2. The results are obtained from 100 MC runs with the EKSPF and the PUESRF.}
\label{fig3}
\end{figure}
\begin{figure}[!t]
\centering
\includegraphics[width=3in]{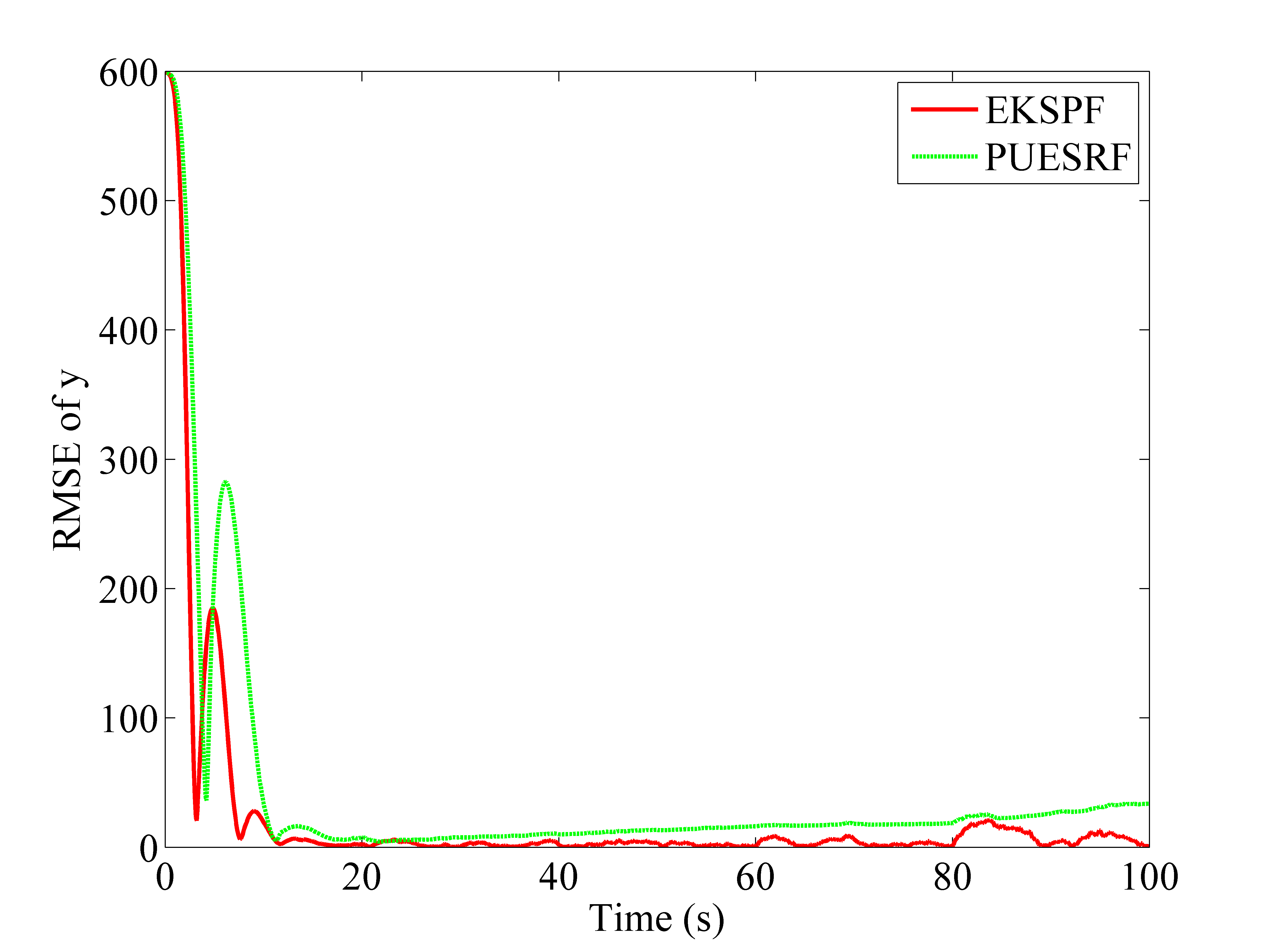}
\caption{RMSE versus time plot of the {\it{y}}-coordinate for the case depicted in Fig. 2. The results are obtained from 100 MC runs with the EKSPF and the PUESRF.}
\label{fig4}
\end{figure}
\begin{figure}[!t]
\centering
\includegraphics[width=3in]{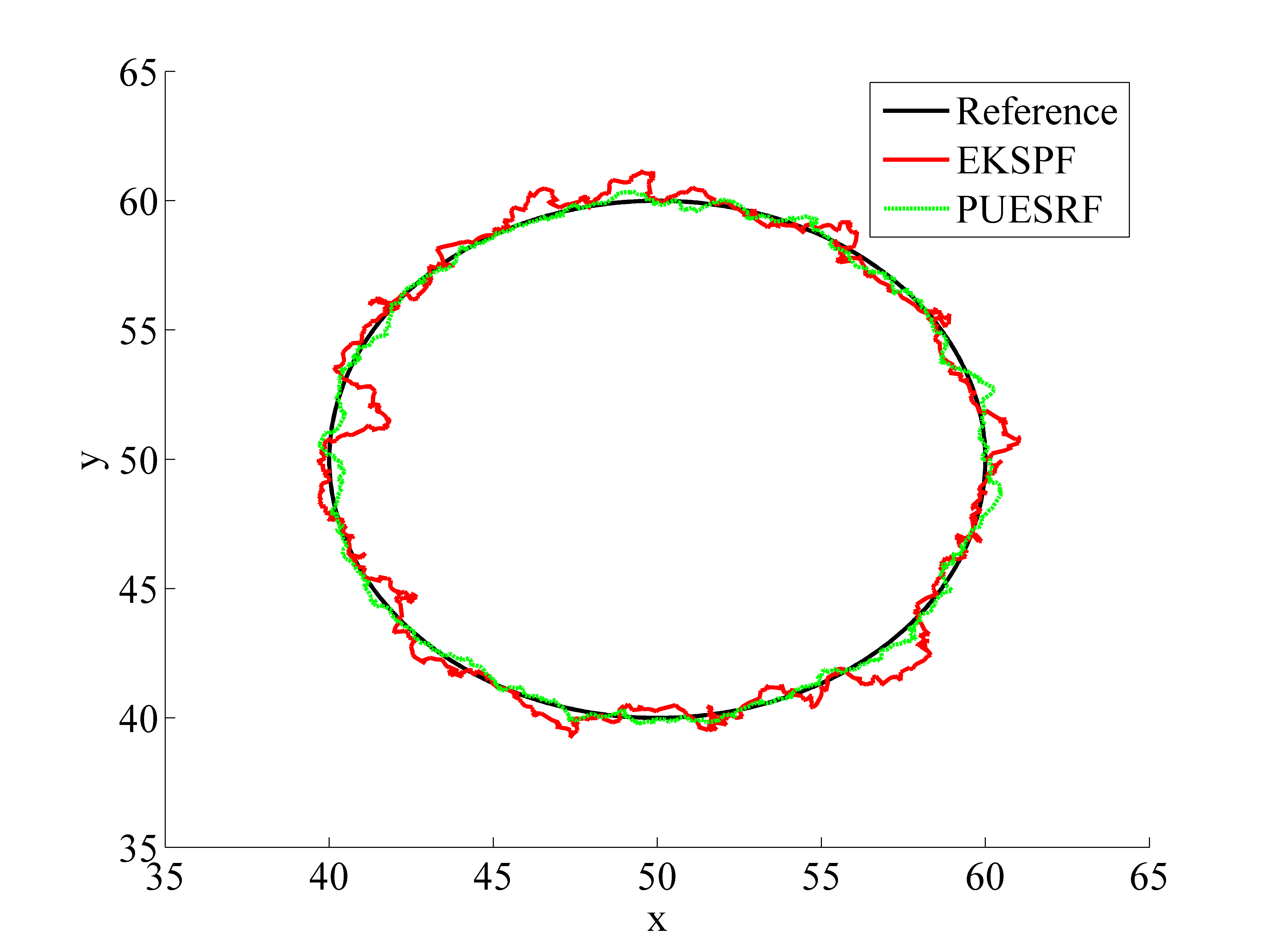}
\caption{Illustration of the filters tracking a circular trajectory.}
\label{fig5}
\end{figure}

\subsection{State and parameter estimation of a mechanical oscillator under dynamic loading}
In this problem, we estimate the states and parameters of a multi-degree of freedom (MDOF) shear frame model. This problem is posed as one with Poisson-type measurements to assess the performance of the filter in such a case. The Poisson measurements are generated in the numerical simulations by allowing its rate to be modulated by a few of the (hidden) states to be estimated. The governing differential equation describing the system model of the MDOF oscillator is given by \[{{\bf{\ddot X}}_t} + {\bf{C}}{{\bf{\dot X}}_t} + {\bf{K}}{{\bf{X}}_t} = {\bf{F}}\left( t \right) + {\bf{\Sigma }}{{\bf{\dot B}}_t}\]
where the viscous damping matrix is of the form \({\bf{C}} = \left[ {\begin{array}{*{20}{c}}{{C^1} + {C^2}}&{ - {C^2}}&0&0&0\\{ - {C^2}}&{{C^2} + {C^3}}&{ - {C^3}}&0&0\\0&{ - {C^3}}&{{C^3} + {C^4}}&{ - {C^4}}&0\\0&0&{ - {C^4}}&{{C^4} + {C^5}}&{ - {C^5}}\\0&0&0&{ - {C^5}}&{{C^5}}\end{array}} \right]\). The stiffness matrix \({\bf{K}}\) is of the same form as \({\bf{C}}\) with \({K^i}\)s replacing \({C^i}\)s. Note that \({C^i}\) and \({K^i}\) denote the damping and stiffness parameters respectively of the \({i^{th}}\) floor and \({\bf{F}}\left( t \right) = \left( {{F^1}\left( t \right),...,{F^5}\left( t \right)} \right)\) where \({F^i}\left( t \right) = {f_0}\cos \left( {\omega t} \right),i = 1,...,5\) is the deterministic force applied at the \({i^{th}}\) floor. Moreover, the diffusion matrix \({\boldsymbol{\Sigma }}\) is diagonal and \({{\bf{B}}_t}\) is a standard Brownian motion. For the present problem, the state/parameter vector to be estimated is given by  \[{\bf{X'}} = {\left[ {{X^1},...,{X^5},{{\dot X}^1},...,{{\dot X}^5},{K^1},...,{K^5},{C^1},...,{C^5}} \right]^T}\] The synthetic Poisson measurements for the EKSPF are generated by constructing the rate function as, \({\boldsymbol{\lambda }} = \left( {{\lambda ^1},...,{\lambda ^5}} \right)\) with \({\lambda ^i} = {\alpha _i}\left| {{X^i}} \right|,i = 1,...,5\) where \({\alpha _i} = {10^6}\). The measurement equation then takes the form of (\ref{eq2}) and is of a purely non-diffusive type. The synthetic measurements for the PUESRF are however generated by adding 1\% standard Gaussian noise to the displacement vector, \({{\bf{X}}_t}\), the solution obtained from the governing differential equation, whilst assuming the stiffness/damping parameters to be known. Such values of the various parameters (taken as the reference) used for the simulations are as follows: \({f_0} = 30N\), \(\omega  = 1rad/s\), \({C^i} = 5Ns/m\), \({K^i} = 100N/m\) for \(i = 1,...,5\). The diagonal entries in \({\boldsymbol{\Sigma }}\) are 0.01.  
\par The performance of the EKSPF is compared with the PUESRF although both the filters take different forms of measurements as input. While Figs. 6 and 7 show the damping parameter estimates of the EKSPF and the PUESRF, Figs. 8 and 9 give the respective plots for the stiffness parameters. Clearly, the EKSPF provides a better estimation of the parameters in comparison to the PUESRF. This is further highlighted in the RMSE plots. As in the previous example, the RMSE is calculated from 100 MC runs of both filters. For brevity, the plots are shown only for the parameters. 

\par Through this example, we have numerically shown the efficacy of the EKSPF in the recursive estimation of states in multi-dimensional nonlinear systems with Poisson measurements. The proposed filtering scheme may be contrasted with the existing ones employing Poisson-type measurements \cite{snyder}-\cite{mje} wherein numerical demonstrations are limited to low-dimensional filtering problems, e.g. those involving a maximum of only 2 unknowns. We have also numerically demonstrated that the EKSPF is a viable filtering tool for diffusion-type measurements as well. Finally, in the next subsection, we pose a problem of active structural control in a filtering framework. 

\begin{figure}[!t]
\centering
\includegraphics[width=3in]{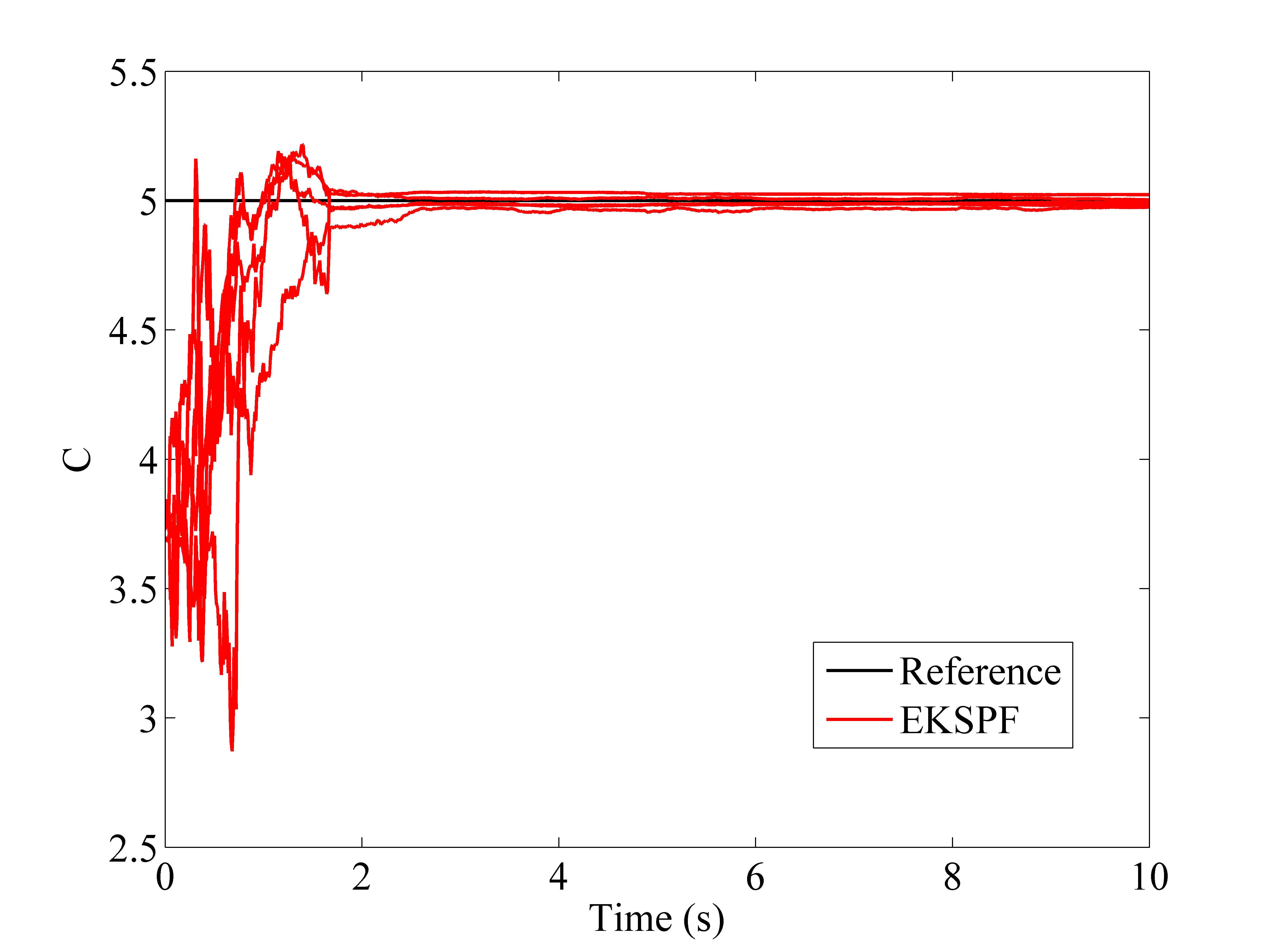}
\caption{EKSPF estimates of the damping parameters.}
\label{fig6}
\end{figure}
\begin{figure}[!t]
\centering
\includegraphics[width=3in]{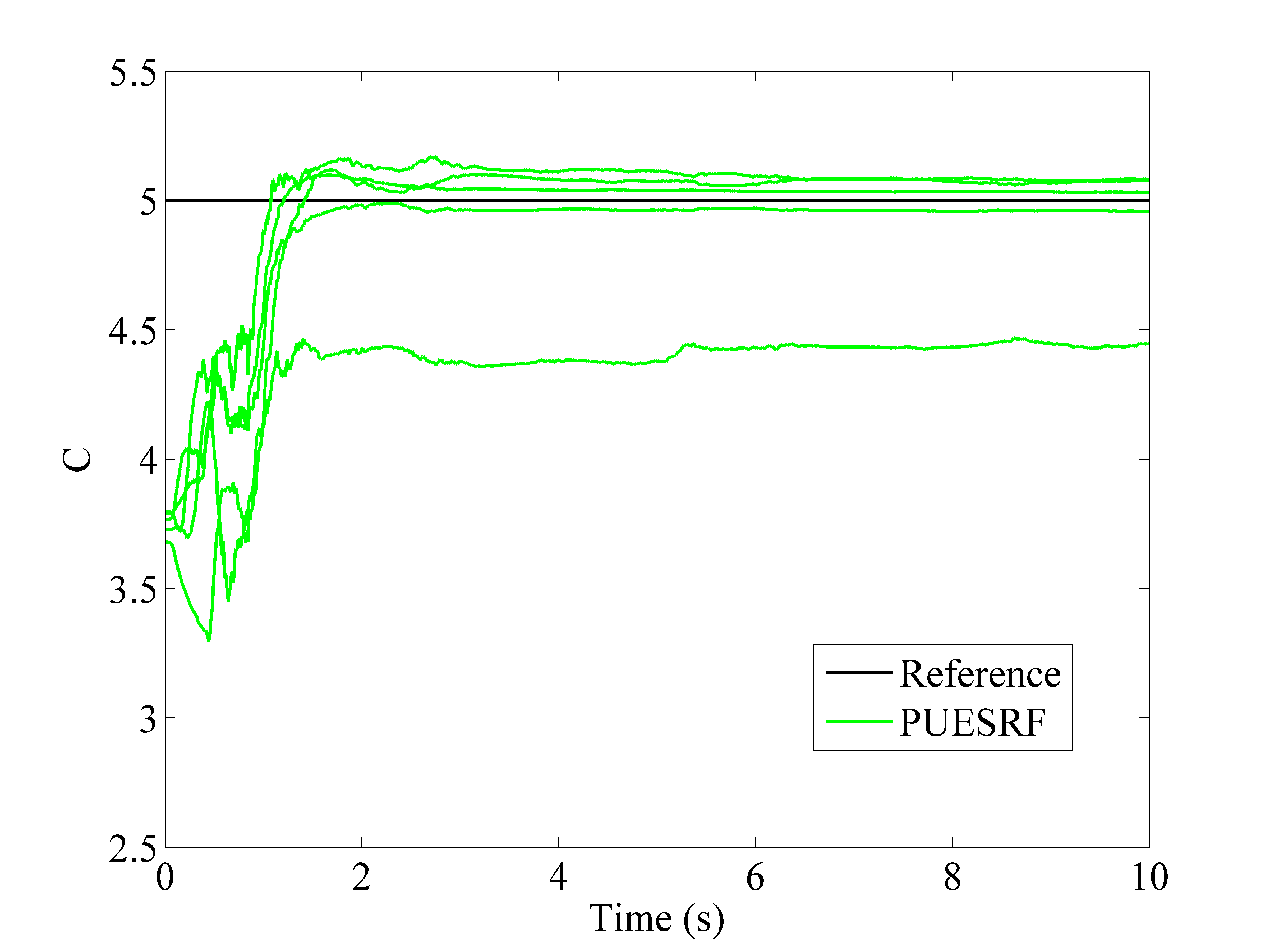}
\caption{PUESRF estimates of the damping parameters.}
\label{fig7}
\end{figure}
\begin{figure}[!t]
\centering
\includegraphics[width=3in]{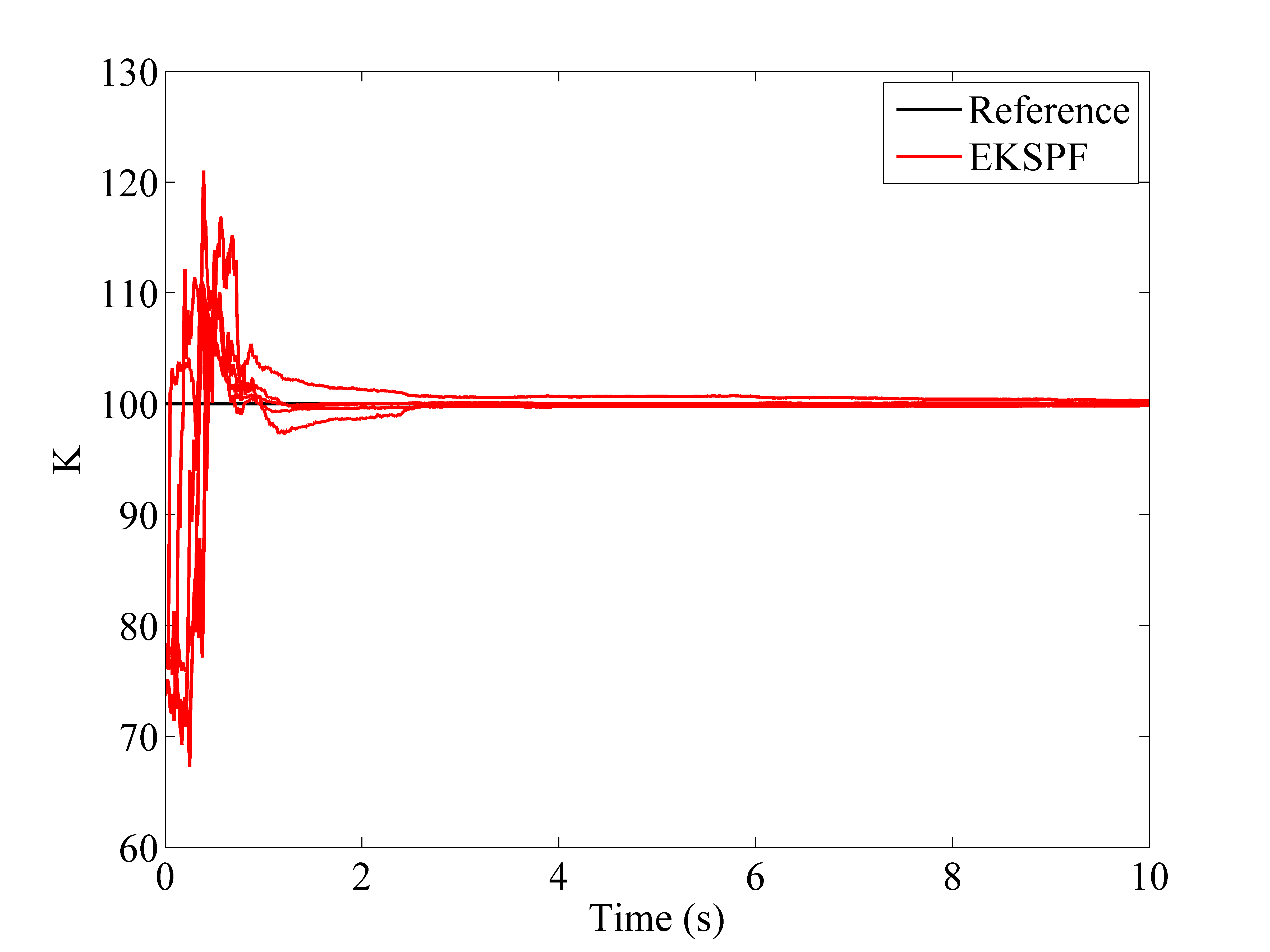}
\caption{EKSPF estimates of the stiffness parameters.}
\label{fig8}
\end{figure}
\begin{figure}[!t]
\centering
\includegraphics[width=3in]{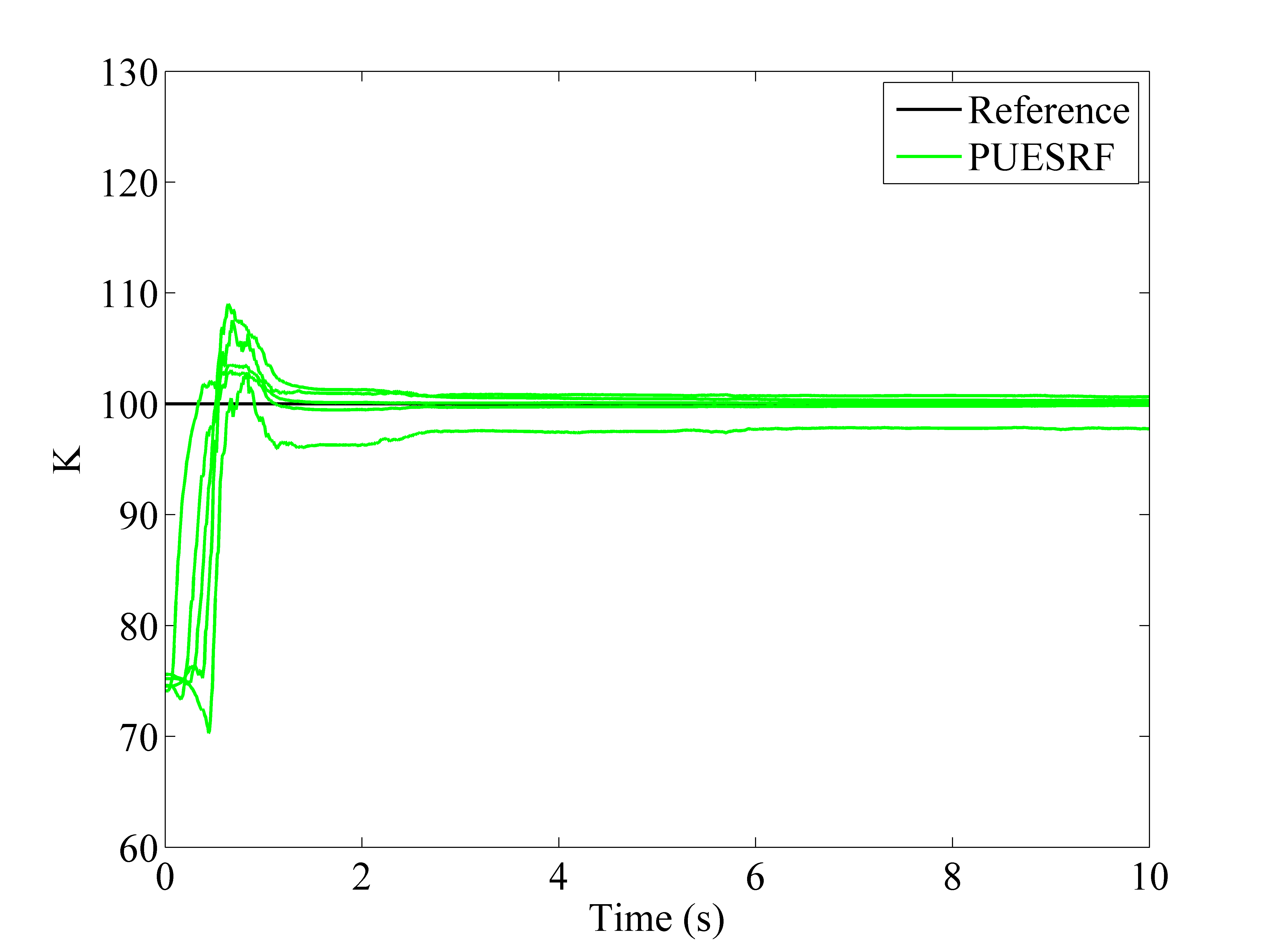}
\caption{PUESRF estimates of the stiffness parameters.}
\label{fig9}
\end{figure}
\begin{figure}[!t]
\centering
\includegraphics[width=3in]{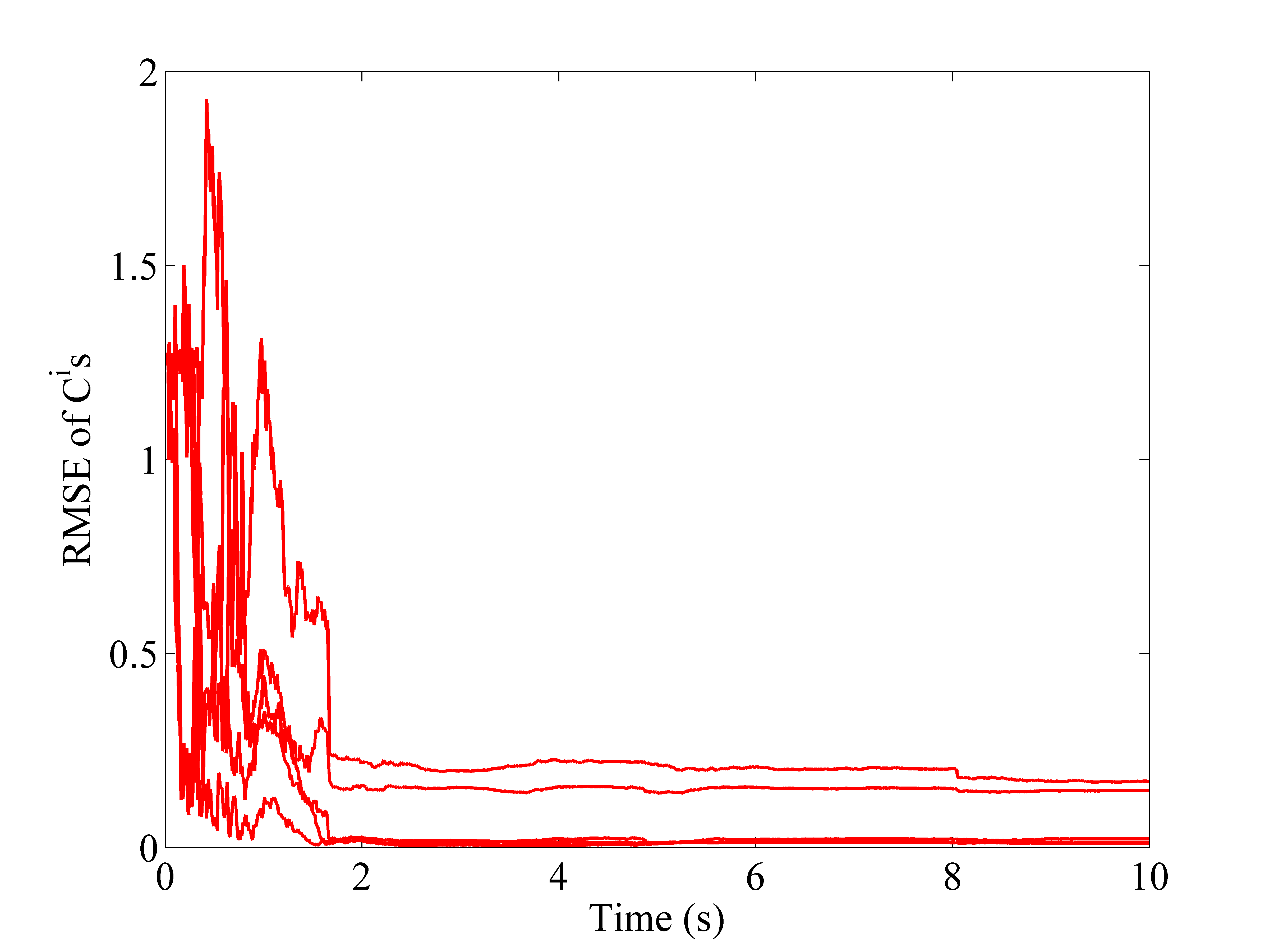}
\caption{RMSE versus time plot of the damping parameters estimated by the EKSPF. The results are obtained from 100 MC runs.}
\label{fig10}
\end{figure}
\begin{figure}[!t]
\centering
\includegraphics[width=3in]{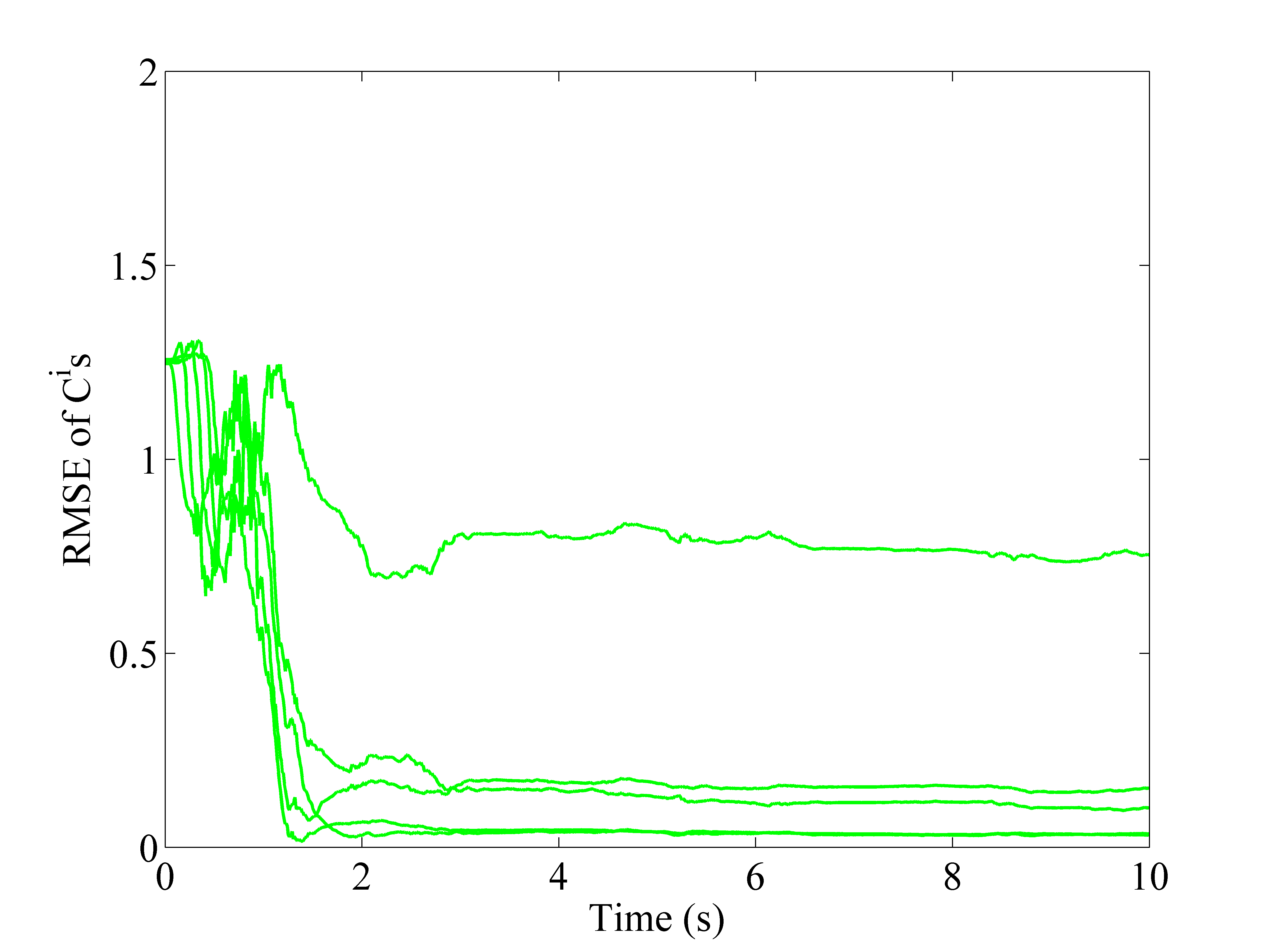}
\caption{RMSE versus time plot of the damping parameters estimated by the PUESRF. The results are obtained from 100 MC runs.}
\label{fig11}
\end{figure}
\begin{figure}[!t]
\centering
\includegraphics[width=3in]{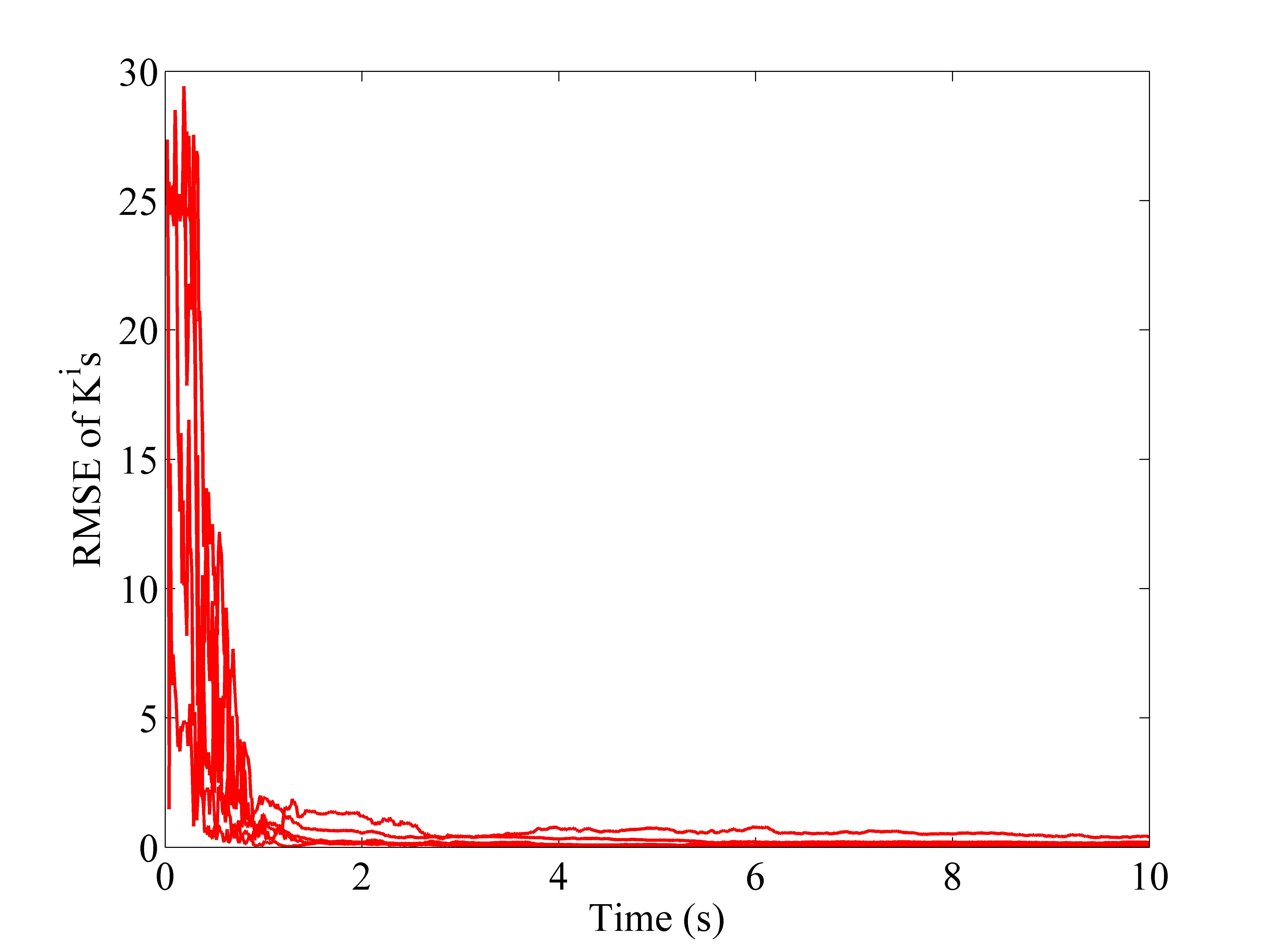}
\caption{RMSE versus time plot of the stiffness parameters estimated by the EKSPF. The results are obtained from 100 MC runs.}
\label{fig12}
\end{figure}
\begin{figure}[!t]
\centering
\includegraphics[width=3in]{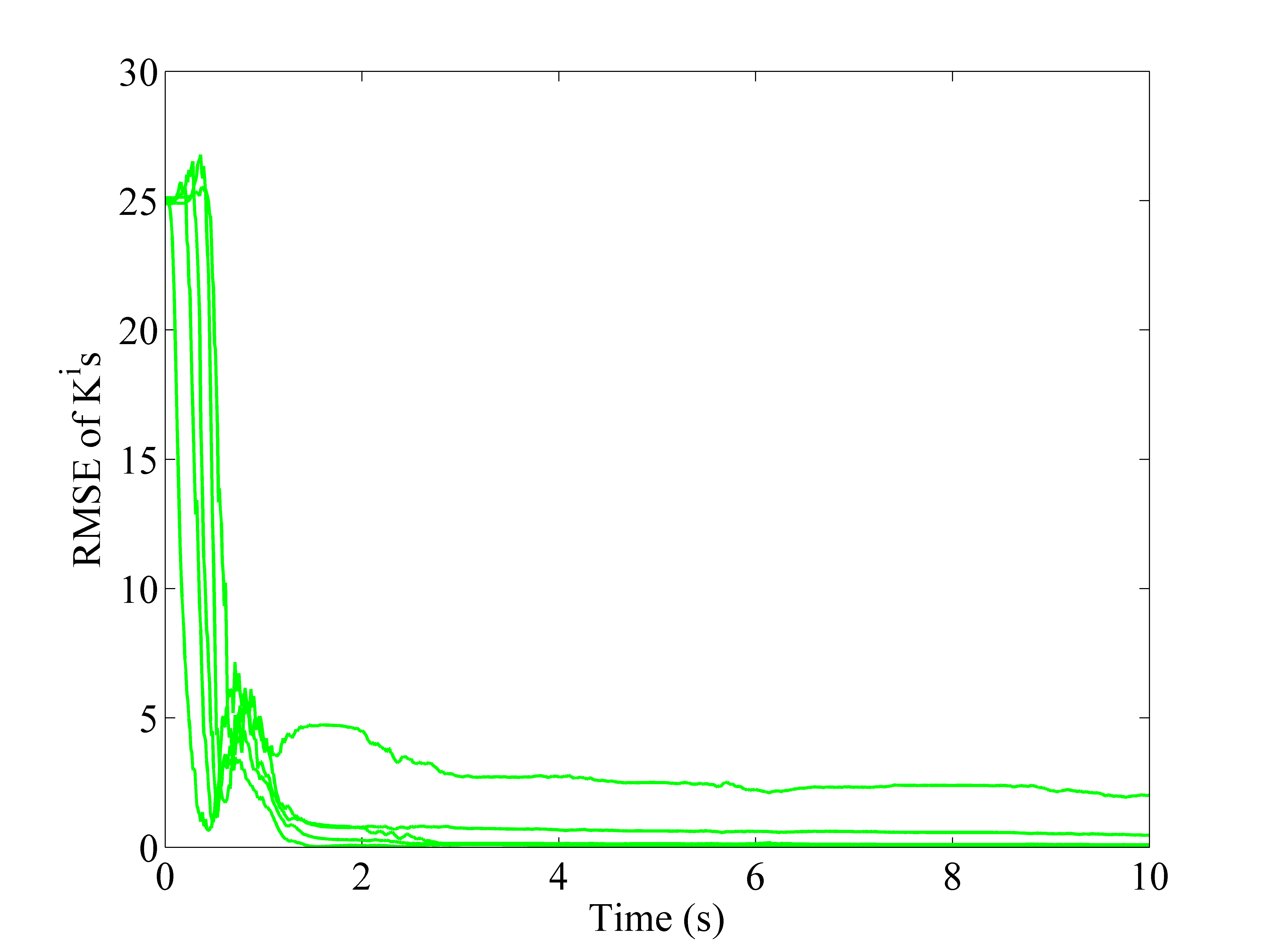}
\caption{RMSE versus time plot of the stiffness parameters estimated by the PUESRF. The results are obtained from 100 MC runs.}
\label{fig13}
\end{figure}

\subsection{Active structural control of a mechanical oscillator}
Many engineering structures must be so designed as to withstand the ever-changing dynamical consequences of the environment such as the winds, earthquakes etc. In such cases, the passive control elements alone might prove insufficient to restrict the structural responses within safe limits. This has led to the development of various active control schemes, for both linear and nonlinear systems. Some of these schemes are gain scheduling, sliding mode control, feedback linearization and the state-dependent Riccati equation method (see \cite{soong}-\cite{cloutier} and the references therein). However, as pointed out in \cite{gj}, the mathematical models developed in these schemes result in inadequate control algorithms when the uncertainties (such as measurement noises) involved are considerable. This is because these schemes are programmed to carry out specific control tasks based on the system analysis. On the other hand, a neural network based control algorithm can learn to adapt to the changing dynamical patterns on the go. Nevertheless, research is still under way in developing neurocontroller-based schemes as the initial training procedure is not straightforward. We pose the problem of active control within a filtering framework wherein the goal is to estimate the control force that extremizes an appropriately chosen performance index. The details pertaining to such a characterization are given next along with the accompanying equations. The (nonlinear) problem chosen for the present study is to minimize the structural response of a duffing oscillator under external forcing. 

\par The governing differential equation of the nonlinear (Duffing) oscillator, subjected to base acceleration and external forcing is given by \[\ddot X\left( t \right) + C\dot X\left( t \right) + KX\left( t \right) + D{X^3}\left( t \right) = F\left( t \right) + \Sigma \dot B - {\ddot X_g}\left( t \right) + U\left( t \right)\]
where $D$ denotes a scalar value, \({\ddot X_g}\) the base acceleration and \({U_t}: = U\left( t \right)\) the control force that needs to be estimated. Note that the remaining symbols appearing in the above equation denote the states/parameters similar to the MDOF problem of the previous subsection and that the non-bold letters signify scalar variables. In the present study, the performance index considered below is of a typical form employed in instantaneous control algorithms \cite{soong}  \[{J_t} = {{\bf{X'}}_t}^T{\bf{R}}{{\bf{X'}}_t} + {U_t}^T{\bf{S}}{U_t}\]
where \({\bf{X'}} = {\left[ {X\,\dot X} \right]^T}\) and \({\bf{R}},{\bf{S}}\) are weighting matrices which are assigned according to the relative importance attached to the states and the control forces in the minimization. The aim here is to estimate the control force, \({U_t}\) so as to minimize \({J_t}\) at each time instant. The minimization may be attempted in a filtering framework by posing the unknown control force as the state vector to be estimated by conditioning on the available minimum of the performance index providing the measurement. Subject to the mean-square integrability of all the stochastic processes involved, this is equivalent to projecting the control force on the available minimum of the performance index. In such a characterization, the only missing link is a precise form of the `measurement innovation' that is naturally available in a conventional filtering problem. Here, considering the MC set-up employed in our numerical scheme, we take \({M_t}-{J_t}\) as the measurement innovation term, where \({M_t}= \min \left( {{J_t}\left( {\tilde U\left( j \right)} \right),j = 1,...,{n_e}} \right)\) denotes the available empirical minimum of $J_t$ at time \(t\). Thus, at any given time instant, the update step of filtering drives all the particles so as to render the above innovation a zero-mean martingale. However given the nonlinearity of the innovation term in the control force $U_t$ and the ensemble being a necessarily finite representation of the sample space, an update procedure at any given time should ideally be iterative. Accordingly, at time \(t\), we introduce inner iterations, the aim of which is to iteratively guide the updates so as to impart to $M_t-J_t$ the structure of a zero-mean Poisson martingale as a function of the iterations. Note that the problem of functional minimization has been considered as a Brownian martingale problem in a recent article by the third author and co-workers \cite{psdrr}.

\par Having defined the filtering scheme as adapted to the control problem and treating the `measurement innovation' to be of the diffusion-type, the problem can be solved using {\it{Algorithm 2}}, presently with \(\alpha  = {10^4}\). The various parameter values are as follows: \(C = 5Ns/m\), \(K = 100N/m\), \({f_0} = 20N\), \(\omega  = 5rad/s\) , \(\Sigma  = 0.01\) and \({X_g}\left( t \right) = {X_{g0}}\sin \left( {{\omega _g}t} \right)\) with \({X_{g0}} = 0.05m\) and \({\omega _g} = 4rad/s\). In the absence of a process model, we use the following prediction equation for \({U_t}\), \[{\tilde U_{{t_{i + 1}}}}\left( j \right) = {\hat U_{{t_i}}}\left( j \right) + \Sigma \Delta B\left( j \right),j = 1,...,{n_e}\] where \(\Sigma\) is taken as 50. In the present problem, the inner iterations involving steps 4-6 of {\it{Algorithm 2}} are repeated 5 times at every time step, such that at the \({i^{th}}\) inner iteration, the predicted values are given by the \({\left( {i - 1} \right)^{th}}\) update values. In the first iteration, however, the predicted solution at the current time is made use of. 

\par While Figs. 14 and 15 show the time histories of the uncontrolled and controlled displacements and velocities respectively of the oscillator, Fig. 16 plots the estimated control force against the external force applied. The controlled nature of the system response is evident in these figures. 

\begin{figure}[!t]
\centering
\includegraphics[width=3in]{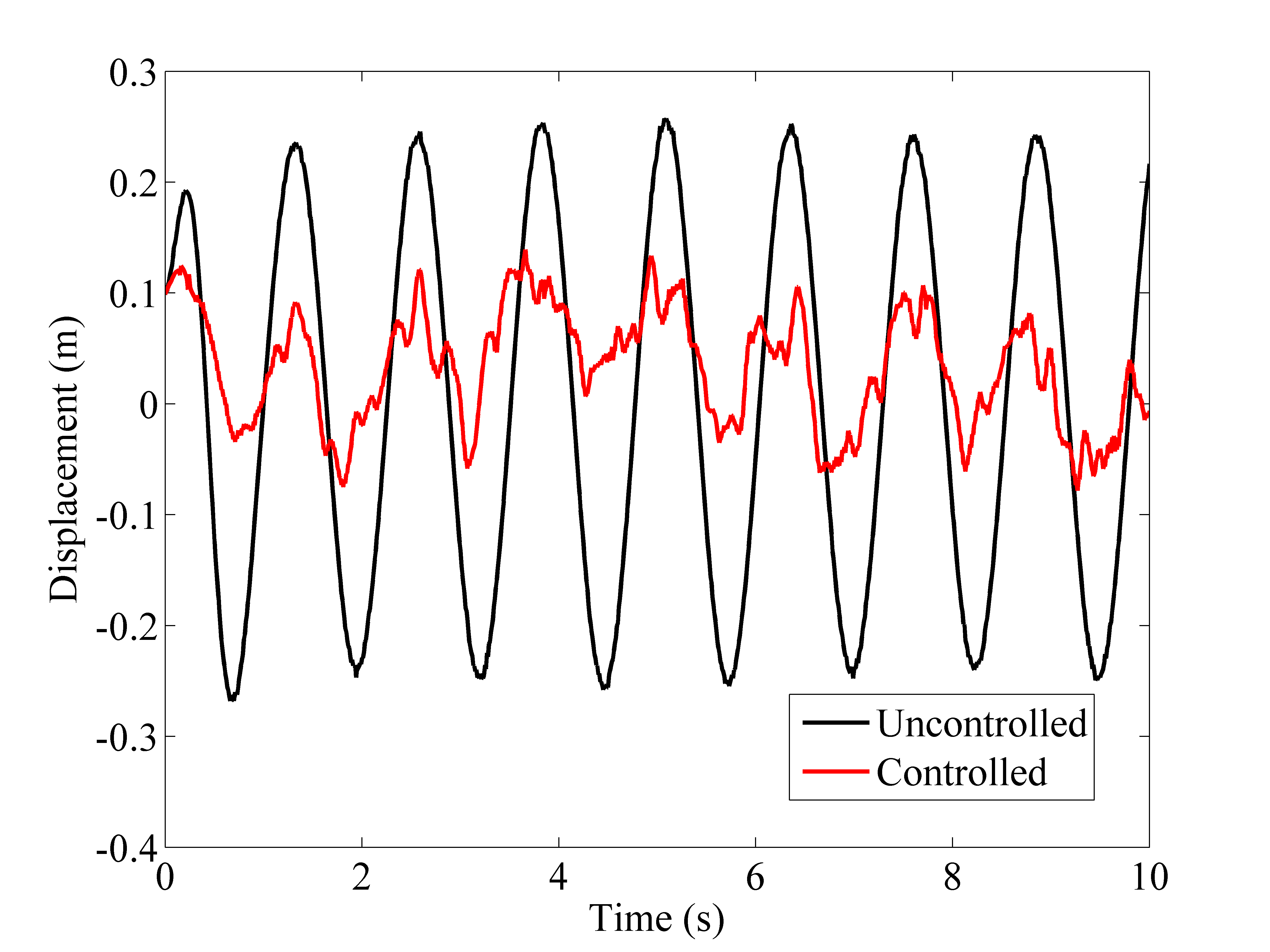}
\caption{A plot of the time histories of the uncontrolled and controlled displacements.}
\label{fig14}
\end{figure}
\begin{figure}[!t]
\centering
\includegraphics[width=3in]{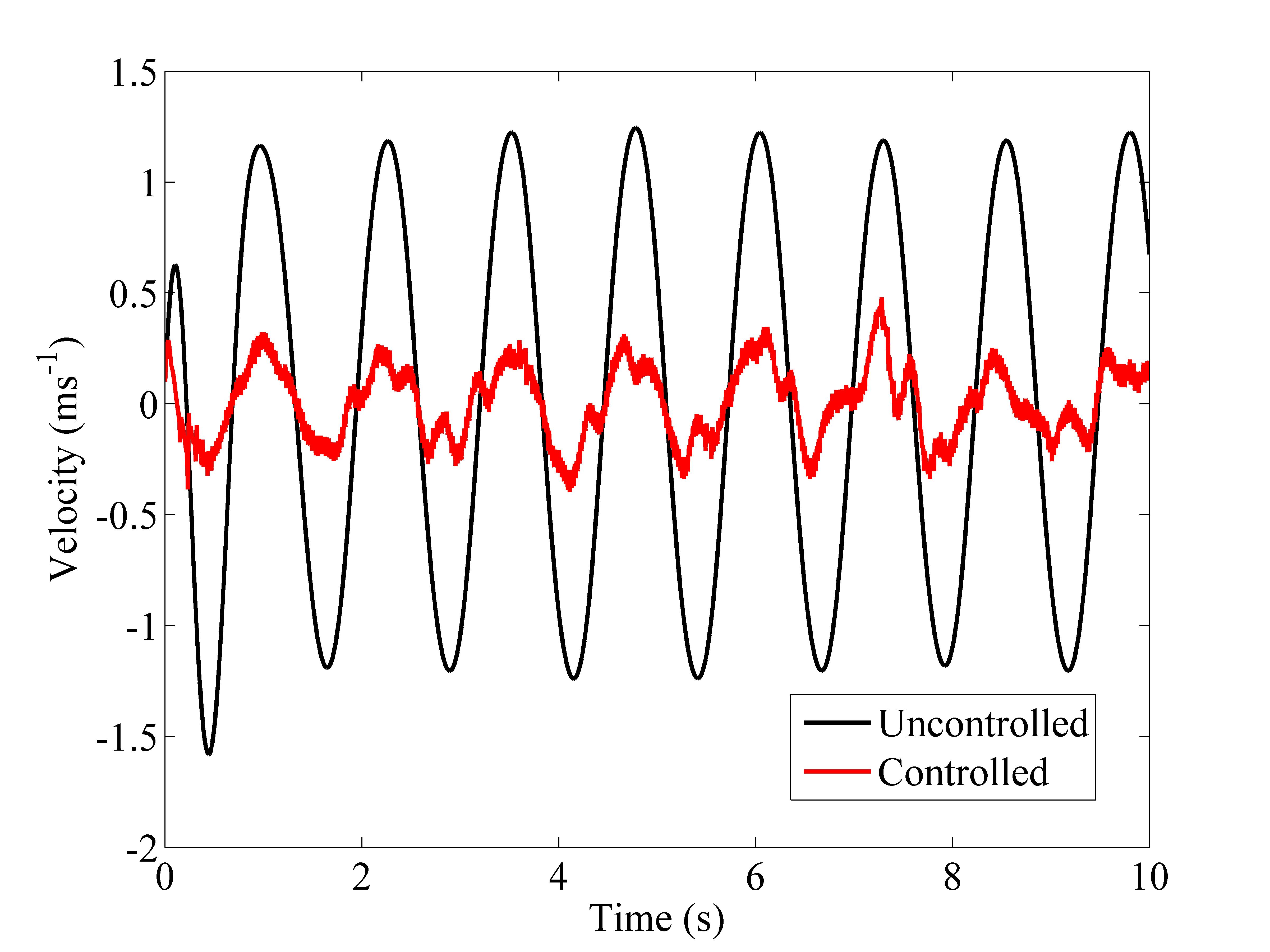}
\caption{A plot of the time histories of the uncontrolled and controlled velocities.}
\label{fig15}
\end{figure}
\begin{figure}[!t]
\centering
\includegraphics[width=3in]{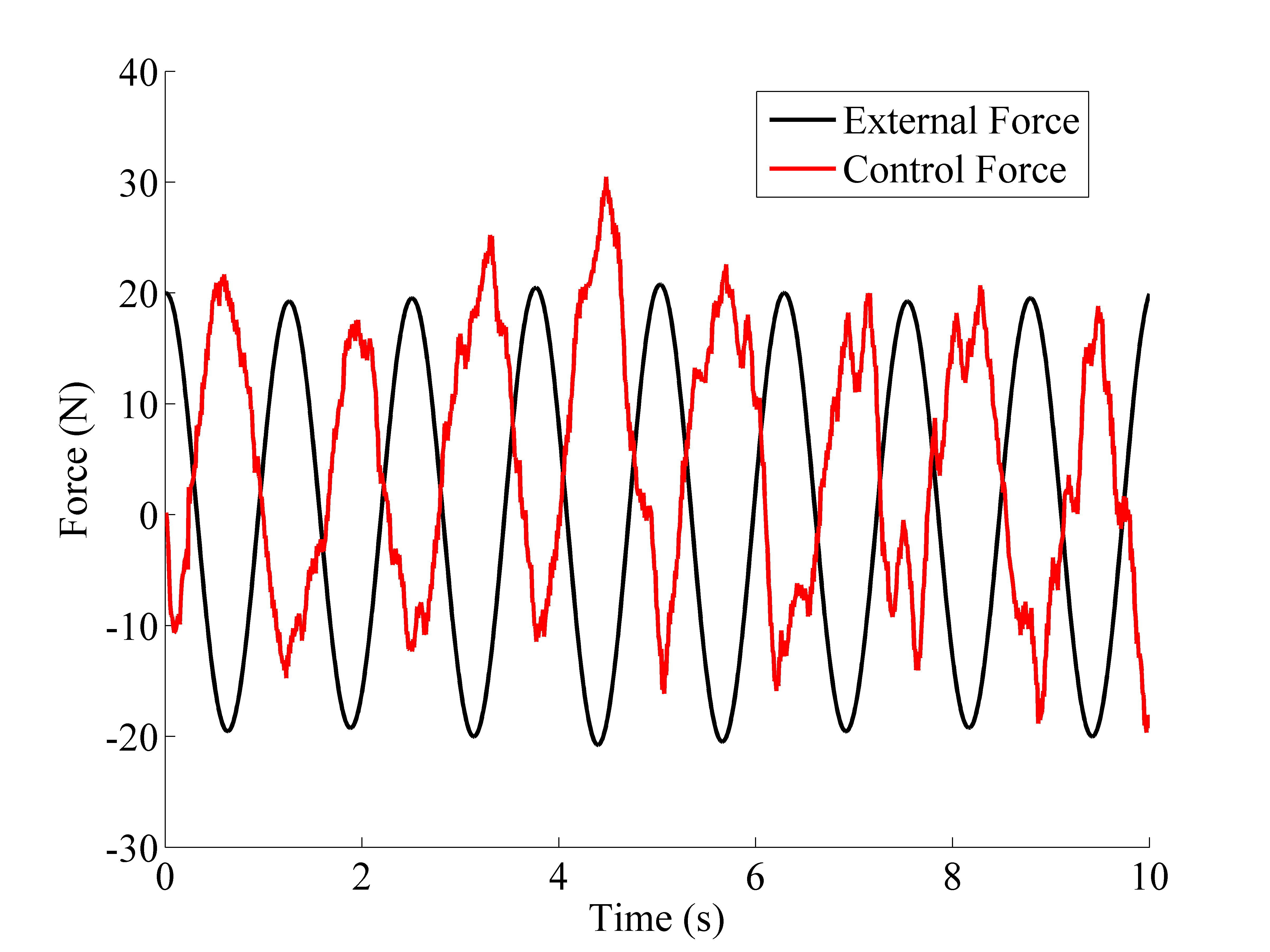}
\caption{A plot of the time histories of the external force and the estimated control force.}
\label{fig16}
\end{figure}

\section{Conclusion}
A novel proposal for Monte Carlo filtering, applicable to state/parameter estimation of nonlinear dynamical systems with Poisson measurement noises, is set forth in this study. Motivated by the form of the governing nonlinear filtering equation of the integro-differential form, herein referred to as the Kushner-Stratonovich-Poisson (KSP) equation, a gain-like additive update is applied to the predicted particles. The update is derived based on a change of measures that aim at rendering the measurement-prediction mismatch, also called the innovation, a zero-mean Poisson martingale. The additive nature of the update, in lieu of the weight-based multiplicative updates employed with many Monte Carlo schemes, ameliorates the problem of particle impoverishment and hence enables applying the present scheme to higher dimensional filtering problems. As demonstrated in this work, the proposed approach admits a ready extension to filtering problems involving purely diffusive measurement noise and, by appropriately redefining the innovation term, even to automatic control problems that demand an estimation of the active control force.

\appendices
\section{Proof of the KSP Equation}
From {\it{Lemma 1}}, we have \[d{\sigma _t}\left( \phi  \right) = {\sigma _t}\left( {L\phi } \right)dt + {\sigma _{t - }}\left( {\phi \left( {{\lambda _t} - 1} \right)} \right)\left( {d{Y_t} - dt} \right)\] and \[\begin{array}{l}{\sigma _t}\left( 1 \right) = 1 + \int\limits_0^t {{\sigma _{s - }}\left( {{\lambda _s} - 1} \right)\left( {d{Y_s} - ds} \right)} \\\,\,\,\,\,\,\,\,\,\,\,\,\,\,\,\, = 1 + \int\limits_0^t {{\sigma _s}\left( 1 \right)\left( {{\pi _{s - }}\left( \lambda  \right) - 1} \right)\left( {d{Y_s} - ds} \right)} \,\,\,\end{array}\]
Now denote \({\sigma _t}\left( 1 \right) = {{\rm{E}}_{\rm{Q}}}\left[ {{Z_t}|{\mathcal{F}}_t^y} \right]\) by \({\mathord{\buildrel{\lower3pt\hbox{$\scriptscriptstyle\frown$}} \over Z} _t}\). Then, we have \[\begin{array}{l}{{\mathord{\buildrel{\lower3pt\hbox{$\scriptscriptstyle\frown$}} \over Z} }_t} = {{\rm{E}}_{\rm{Q}}}\left[ {1 + \int\limits_0^t {{Z_{s - }}\left( {{\lambda _s} - 1} \right)\left( {d{Y_s} - ds} \right)} |{\mathcal{F}}_t^y} \right]\\\,\,\,\,\,\,\,\, = 1 + \int\limits_0^t {{{\mathord{\buildrel{\lower3pt\hbox{$\scriptscriptstyle\frown$}} \over Z} }_{s - }}\left( {{{\mathord{\buildrel{\lower3pt\hbox{$\scriptscriptstyle\frown$}} \over \lambda } }_s} - 1} \right)\left( {d{Y_s} - ds} \right)} \end{array}\]
where \({\mathord{\buildrel{\lower3pt\hbox{$\scriptscriptstyle\frown$}} \over \lambda } _s}: = {{\rm{E}}_{\rm{Q}}}\left[ {{\lambda _s}|{\mathcal{F}}_s^y} \right]\). Comparing this expression with the one as above for \({\sigma _t}\left( 1 \right)\), we obtain \({\mathord{\buildrel{\lower3pt\hbox{$\scriptscriptstyle\frown$}} \over \lambda } _t} = {\pi _{t - }}\left( \lambda  \right)\). Since, \({\pi _t}\left( \phi  \right) = {\sigma _t}\left( \phi  \right){\left( {{\sigma _t}\left( 1 \right)} \right)^{ - 1}}\), the next step is to find \(\mathord{\buildrel{\lower3pt\hbox{$\scriptscriptstyle\frown$}} \over Z} _t^{ - 1}\).  Towards this, we recall It\^{o}'s formula for semimartingales \cite{klebaner}, which, for a semimartingale, \({V_t}: = V\left( t \right)\) and a \({\mathbb{C}^2}\) function \(f\) may be given as 
\begin{IEEEeqnarray}{c}
f\left( {{V_t}} \right) = f\left( {{V_0}} \right) + \int\limits_0^t {f'\left( {{V_{s - }}} \right)d{V_s}}  + \frac{1}{2}\int\limits_0^t {f''\left( {{V_{s - }}} \right)d{{\left[ {{V_s},{V_s}} \right]}^c}} \IEEEnonumber* \\ 
 + \sum\limits_{s \leqslant t} {\left( {f\left( {{V_s}} \right) - f\left( {{V_{s - }}} \right) - f'\left( {{V_{s - }}} \right)\Delta {V_s}} \right)} \IEEEnonumber*
\end{IEEEeqnarray}
Now, suppose that a jump occurs at time \(t\); then we have \({\mathord{\buildrel{\lower3pt\hbox{$\scriptscriptstyle\frown$}} \over Z} _t} = {\mathord{\buildrel{\lower3pt\hbox{$\scriptscriptstyle\frown$}} \over \lambda } _t}{\mathord{\buildrel{\lower3pt\hbox{$\scriptscriptstyle\frown$}} \over Z} _{t - }}\). Hence, for any time $t$, \({\mathord{\buildrel{\lower3pt\hbox{$\scriptscriptstyle\frown$}} \over Z} _t} = {\mathord{\buildrel{\lower3pt\hbox{$\scriptscriptstyle\frown$}} \over \lambda } _t}{\mathord{\buildrel{\lower3pt\hbox{$\scriptscriptstyle\frown$}} \over Z} _{t - }}\Delta {Y_t}\). Then, \[f\left( {{V_s}} \right) - f\left( {{V_{s - }}} \right) = \frac{1}{{{{\mathord{\buildrel{\lower3pt\hbox{$\scriptscriptstyle\frown$}} \over Z} }_s}}} - \frac{1}{{{{\mathord{\buildrel{\lower3pt\hbox{$\scriptscriptstyle\frown$}} \over Z} }_{s - }}}} =  - \frac{{{{\mathord{\buildrel{\lower3pt\hbox{$\scriptscriptstyle\frown$}} \over \lambda } }_s} - 1}}{{{{\mathord{\buildrel{\lower3pt\hbox{$\scriptscriptstyle\frown$}} \over \lambda } }_s}{{\mathord{\buildrel{\lower3pt\hbox{$\scriptscriptstyle\frown$}} \over Z} }_{s - }}}}\Delta {Y_s}\]
and \[f'\left( {{V_{s - }}} \right)\Delta {V_s} =  - \frac{1}{{\mathord{\buildrel{\lower3pt\hbox{$\scriptscriptstyle\frown$}} \over Z} _{s - }^2}}\Delta {\mathord{\buildrel{\lower3pt\hbox{$\scriptscriptstyle\frown$}} \over Z} _s} =  - \frac{{{{\mathord{\buildrel{\lower3pt\hbox{$\scriptscriptstyle\frown$}} \over \lambda } }_s} - 1}}{{{{\mathord{\buildrel{\lower3pt\hbox{$\scriptscriptstyle\frown$}} \over Z} }_{s - }}}}\Delta {Y_s}\]
This gives
\begin{IEEEeqnarray}{c}
\sum\limits_{s \leqslant t} {\left( {f\left( {{V_s}} \right) - f\left( {{V_{s - }}} \right) - f'\left( {{V_{s - }}} \right)\Delta {V_s}} \right)} \IEEEnonumber*\\
 =  - \int\limits_0^t {\frac{{{{\overset{\lower0.5em\hbox{$\smash{\scriptscriptstyle\frown}$}}{\lambda } }_s} - 1}}{{{{\overset{\lower0.5em\hbox{$\smash{\scriptscriptstyle\frown}$}}{\lambda } }_s}{{\overset{\lower0.5em\hbox{$\smash{\scriptscriptstyle\frown}$}}{Z} }_{s - }}}}d{Y_s}}  + \int\limits_0^t {\frac{{{{\overset{\lower0.5em\hbox{$\smash{\scriptscriptstyle\frown}$}}{\lambda } }_s} - 1}}{{{{\overset{\lower0.5em\hbox{$\smash{\scriptscriptstyle\frown}$}}{Z} }_{s - }}}}d{Y_s}} \IEEEnonumber*
\end{IEEEeqnarray}
Also, \[\int\limits_0^t {f'\left( {{V_{s - }}} \right)d{V_s}}  = \int\limits_0^t { - \frac{1}{{\mathord{\buildrel{\lower3pt\hbox{$\scriptscriptstyle\frown$}} \over Z} _{s - }^2}}d{{\mathord{\buildrel{\lower3pt\hbox{$\scriptscriptstyle\frown$}} \over Z} }_s}}  =  - \int\limits_0^t {\frac{{{{\mathord{\buildrel{\lower3pt\hbox{$\scriptscriptstyle\frown$}} \over \lambda } }_s} - 1}}{{{{\mathord{\buildrel{\lower3pt\hbox{$\scriptscriptstyle\frown$}} \over Z} }_{s - }}}}\left( {d{Y_s} - ds} \right)} \]
Now combining all the above terms and noting that \(f\left( {{V_0}} \right) = 1\) and \(\frac{1}{2}\int\limits_0^t {f''\left( {{V_{s - }}} \right)d{{\left[ {{V_s},{V_s}} \right]}^c}}  = 0\), we have \[\mathord{\buildrel{\lower3pt\hbox{$\scriptscriptstyle\frown$}} \over Z} _t^{ - 1} = 1 - \int\limits_0^t {\frac{{{{\mathord{\buildrel{\lower3pt\hbox{$\scriptscriptstyle\frown$}} \over \lambda } }_s} - 1}}{{{{\mathord{\buildrel{\lower3pt\hbox{$\scriptscriptstyle\frown$}} \over \lambda } }_s}{{\mathord{\buildrel{\lower3pt\hbox{$\scriptscriptstyle\frown$}} \over Z} }_{s - }}}}\left( {d{Y_s} - {{\mathord{\buildrel{\lower3pt\hbox{$\scriptscriptstyle\frown$}} \over \lambda } }_s}ds} \right)} \]
Again, by It\^{o}'s formula, \[\begin{array}{l}d{\pi _t}\left( \phi  \right) = {\sigma _t}\left( \phi  \right)d\left( {\mathord{\buildrel{\lower3pt\hbox{$\scriptscriptstyle\frown$}} \over {\rm Z}} _t^{ - 1}} \right) + \mathord{\buildrel{\lower3pt\hbox{$\scriptscriptstyle\frown$}} \over {\rm Z}} _t^{ - 1}d{\sigma _t}\left( \phi  \right) + \sum\limits_{s \le t} {\Delta {\sigma _s}\left( \phi  \right)\Delta \mathord{\buildrel{\lower3pt\hbox{$\scriptscriptstyle\frown$}} \over {\rm Z}} _s^{ - 1}} \\\,\,\,\,\,\,\,\,\,\,\,\,\,\,\,\,\,\,\,\, = {\sigma _t}\left( \phi  \right)\left( { - \frac{{{{\mathord{\buildrel{\lower3pt\hbox{$\scriptscriptstyle\frown$}} \over \lambda } }_t} - 1}}{{{{\mathord{\buildrel{\lower3pt\hbox{$\scriptscriptstyle\frown$}} \over \lambda } }_t}{{\mathord{\buildrel{\lower3pt\hbox{$\scriptscriptstyle\frown$}} \over Z} }_{t - }}}}} \right)\left( {d{Y_t} - {{\mathord{\buildrel{\lower3pt\hbox{$\scriptscriptstyle\frown$}} \over \lambda } }_t}dt} \right)\\\,\,\,\,\,\,\,\,\,\,\,\,\,\,\,\,\,\,\,\,\,\,\,\,\, + \mathord{\buildrel{\lower3pt\hbox{$\scriptscriptstyle\frown$}} \over {\rm Z}} _t^{ - 1}\left( {{\sigma _t}\left( {L\phi } \right)dt + {\sigma _{t - }}\left( {\phi \left( {{\lambda _t} - 1} \right)} \right)\left( {d{Y_t} - dt} \right)} \right)\\\,\,\,\,\,\,\,\,\,\,\,\,\,\,\,\,\,\,\,\,\,\,\,\,\, - {\sigma _{t - }}\left( {\phi \left( {{\lambda _t} - 1} \right)} \right)\frac{{{{\mathord{\buildrel{\lower3pt\hbox{$\scriptscriptstyle\frown$}} \over \lambda } }_t} - 1}}{{{{\mathord{\buildrel{\lower3pt\hbox{$\scriptscriptstyle\frown$}} \over \lambda } }_t}{{\mathord{\buildrel{\lower3pt\hbox{$\scriptscriptstyle\frown$}} \over Z} }_{t - }}}}d{Y_t}\end{array}\]
Using \({\mathord{\buildrel{\lower3pt\hbox{$\scriptscriptstyle\frown$}} \over \lambda } _t} = {\pi _{t - }}\left( \lambda  \right)\) and \({\pi _t}\left( . \right) = {\sigma _t}\left( . \right)\mathord{\buildrel{\lower3pt\hbox{$\scriptscriptstyle\frown$}} \over Z} _t^{ - 1}\) and through simple rearrangements, we obtain the KSP equation
\begin{IEEEeqnarray}{c}
{\pi _t}\left( \phi  \right) = {\pi _0}\left( \phi  \right) + \int\limits_0^t {{\pi _s}\left( {L\phi } \right)ds} + \IEEEnonumber*\\  
\int\limits_0^t {\left\{ {\frac{{{\pi _{s - }}\left( {\lambda \phi } \right) - {\pi _{s - }}\left( \phi  \right){\pi _{s - }}\left( \lambda  \right)}}{{{\pi _{s - }}\left( \lambda  \right)}}} \right\}\left( {d{Y_s} - {\pi _{s - }}\left( \lambda  \right)ds} \right)} \IEEEnonumber*
\end{IEEEeqnarray}
Now, for a general case where \({{\bf{Y}}_t} = \left( {Y_t^1,...,Y_t^{{n_y}}} \right) \in {\mathbb{R}^{{n_y}}}\), the aim of filtering is to drive \(\left\{ {Y_t^i - \int\limits_0^t {\lambda _s^ids} } \right\},i = 1,...,{n_y}\) to Poisson-martingales. Accordingly, we have the Radon-Nikodym derivatives, \[Z_t^i = \exp \left( {\int\limits_0^t {\ln \lambda _s^idY_s^i}  - \int\limits_0^t {\left( {\lambda _s^i - 1} \right)ds} } \right),i = 1,...,{n_y}\] and the corresponding change of measures that ensure that each of the component innovation processes are driven to Poisson martingales under the original measure \({\rm{P}}\). Since it is cumbersome to write an integral expression for the combined Radon-Nikodym derivative, we extend the KSP equation in the one-dimensional case to account for the general case of \(n_y\) dimensions. In that case, the gain matrix, after discretization and ensemble approximation, has \(n_y\) columns instead of one.

\ifCLASSOPTIONcaptionsoff
  \newpage
\fi




\end{document}